\documentclass{article}
\usepackage{arxiv}
\usepackage{cite}
\usepackage[utf8]{inputenc}
\usepackage{amsmath,amssymb} %
\usepackage{xcolor}
\usepackage{float}
\usepackage{algorithm2e}
\usepackage{algorithmic}
\usepackage{subcaption}
\usepackage{csquotes}
\usepackage{url}
\usepackage{array}
\usepackage[thinlines]{easytable}
\usepackage{appendix}
\usepackage{hyperref} 
\usepackage{booktabs}
\newcolumntype{L}{>{\centering\arraybackslash}m{4cm}}
\usepackage{lipsum}
\usepackage{graphicx}
\graphicspath{{./Figures/}{../Figures}}
\RestyleAlgo{ruled}

\title{Pattern Characterization Using Topological Data Analysis: Application to Piezo Vibration Striking Treatment}

\author{
    \textbf{Max M. Chumley}\\
    chumleym@msu.edu\\
    Michigan State University\\
    \and
    \textbf{Melih C. Yesilli}\\
    yesillim@msu.edu\\
    Michigan State University\\
    \and
    \textbf{Jisheng Chen}\\
    chenjish@msu.edu\\
    Michigan State University\\
    \and
    \textbf{Firas A.~Khasawneh$^*$}\\
    khasawn3@msu.edu\\
    Michigan State University\\
    \and
    \textbf{Yang Guo}\\
    yguo@msu.edu\\
    Michigan State University\\
}

\date{}

\begin{document} 

\maketitle
*Address all correspondence to this author.
\section{Abstract}

Quantifying patterns in visual or tactile textures provides important information about the process or phenomena that generated these patterns. In manufacturing, these patterns can be intentionally introduced as a design feature, or they can be a byproduct of a specific process. Since surface texture has significant impact on the mechanical properties and the longevity of the workpiece, it is important to develop tools for quantifying surface patterns and, when applicable, comparing them to their nominal counterparts. While existing tools may be able to indicate the existence of a pattern, they typically do not provide more information about the pattern structure, or how much it deviates from a nominal pattern. Further, prior works do not provide automatic or algorithmic approaches for quantifying other pattern characteristics such as depths' consistency, and variations in the pattern motifs at different level sets. This paper leverages persistent homology from Topological Data Analysis (TDA) to derive noise-robust scores for quantifying motifs' depth and roundness in a pattern. Specifically, sublevel persistence is used to derive scores that quantify the consistency of indentation depths at any level set in Piezo Vibration Striking Treatment (PVST) surfaces. Moreover, we combine sublevel persistence with the distance transform to quantify the consistency of the indentation radii, and to compare them with the nominal ones. Although the tool in our PVST experiments had a semi-spherical profile, we present a generalization of our approach to tools/motifs of arbitrary shapes thus making our method applicable to other pattern-generating manufacturing processes.

\section{Introduction}\label{sec:intro}
Extracting information from surface images is an important field of research with many applications such as medical imaging \cite{Masokano2020}, remote sensing \cite{Ymeti2017,Fan2019}, and metrology. 
In many instances, the texture on the surface represents a pattern with a tessellation of a repeating, base geometric shape called a motif.  
These patterns might be intentionally introduced either for functional reasons, e.g., adding friction, or to realize certain aesthetics. Alternatively, surface patterns can be an inevitable side effect of the process that generated the surface, such as machining marks. 

Characterizing the resulting patterns can provide valuable information on the surface properties, and it can serve as a useful diagnostic of the production process. The quantification of patterns depends on the involved motifs. For example, a pattern of zero-dimensional motifs (points) is characterized by the lattice formed by the points. One-dimensional motifs (lines) can produce patterns that are characterized by the lines' geometry and the spacing between them (for parallel lines). Patterns can also emerge in two dimensions as a result of the line intersections.

Of particular interest is the challenge of characterizing three dimensional patterns imprinted onto nominally planar surfaces. This scenario applies to many scientific domains that use image data to extract information about certain systems or processes. The image can be viewed as a spatial height map that contains information about the motifs. In particular, in this setting quantities of interest include the structure of the two-dimensional projections of the motifs' centroids, the motifs' depths consistency, and the regularity of the shape of the generalized cones produced from intersections of level sets with the motifs. For example, if the motifs are tessellated semi-spheres in the plane, then the quantities of interest are the centers of the circular two-dimensional projections, the depths of the semi-spheres across the surface, and the deviations of the circles' perimeters as a function of the motifs' height. 

One specific field where surface texture description plays an important role is at the intersection of manufacturing and metrology. 
Surface metrology of manufactured parts is directly related to fit, wear, lubrication, and corrosion \cite{Thomas1998} as well as fatigue resistance \cite{Spierings2013,Frazier2014,Chan2012}. 
In additive manufacturing, surface texture is further used to understand and optimize the process \cite{Yin2003,Gu2009,Gu2015}.

In the field of manufacturing, texture analysis is also a valuable quality control tool that can be used to investigate the effectiveness of a manufacturing process and obtain information about the current state of the machine being used \cite{LIU2022108068}. For example, it has been shown that surface textures can be analyzed to identify the occurrence of chatter in a machining process \cite{Khalifa2006,Lei2016,Szydlowski2011,Li2020,Tran2021,Zhu2020}. Surface texture analysis has also been used to monitor and indicate tool wear in a machining process \cite{Bhat2016,Bradley2001,Datta2013,Li2016,Kerr2005,Danesh2015,Kassim2004,Zhu2017}, detect surface defects such as cracks and scratches \cite{stkepien2014research,Santiago2011,xie2008review,ozturk2015comparison,Vijaykumar2015}, and for quantifying surface roughness of a part \cite{Kilic2006, Myshkin2003,Josso2002}. Surface texture can also have a significant effect on the mechanical properties of a part, and as a result, a number of processes have been developed to intentionally introduce surface texture in order to obtain improved mechanical properties. Examples of such a processes include shot peening, elliptical vibration cutting and texturing, and piezo vibration striking treatment (PVST). Shot peening has been shown to improve properties such as the roughness, hardness and wear resistance of a part \cite{AlMangour2016,Hatamleh2008,Liu2020,Maleki2019} and can increase the ultimate and yield strengths \cite{Jamalian2019, Xie2016}. Elliptical vibration cutting is another process that results in a surface texture left behind on the part by inducing another direction of motion in the cutting process creating an elliptical cutting pattern \cite{Guo2013}. These cuts leave a texture behind on the surface of the part that reduces tool wear and burrs, and improved surface properties such as roughness \cite{Kurniawan2016}. Models have also been developed to describe the relationships between the system parameters and the resulting textures for this process \cite{Jiang2020}. Another example of a process that exploits surface texture for improving mechanical properties is piezo vibration striking treatment (PVST) \cite{chen2021force}, see Section~\ref{sec:pvst}. This paper mainly focuses on analyzing results from the PVST process, but avenues are offered for studying textures with differing properties. 

\begin{figure*}[htbp]
    \centering
    \includegraphics[width=\textwidth]{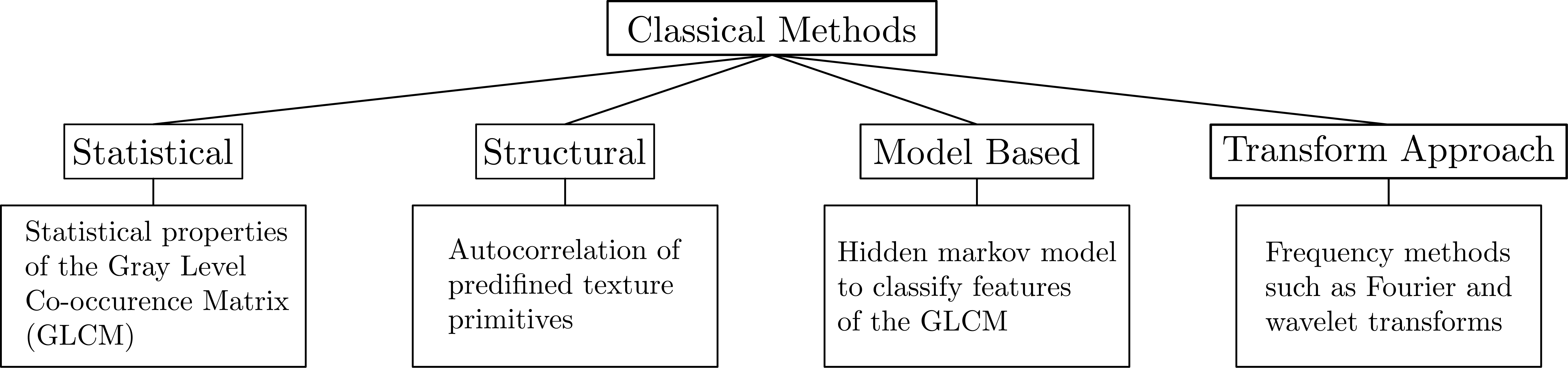}
    \caption{Block diagram summarizing the classical texture analysis methods and their basic descriptions.}
    \label{fig:classical_methods}
\end{figure*}

Most classical applications of texture analysis involve high resolution gray-scale images that provide depth information of the surface. A variety of different methods have been used for analyzing these images ranging from statistical techniques to wavelet transform approaches \cite{Bharati2004, Danesh2015,Josso2002}. The classical approaches can be grouped into four categories that are summarized in Fig.~\ref{fig:classical_methods}.
For statistical methods, the gray level co-occurrence matrix (GLCM) is usually of interest in which a matrix is obtained containing information on the probabilities that adjacent pixels would have the same intensity \cite{Bharati2004}. Statistical measures are then computed on this matrix leading to quantification of broad features such as \textit{smoothness}, \textit{coarseness} and \textit{regularity} of a texture \cite{srinivasan2008statistical,Khalifa2006}. 

Another method of texture analysis is referred to as \textit{structural texture analysis}. This method works best for tessellated patterns of predefined fundamental features called primitives \cite{srinivasan2008statistical}. Statistical quantities such as the image autocorrelation function provide information about the sizing of the primitives and a quantification of the texture periodicity \cite{srinivasan2008statistical}. The problem with this method is that the primitives and relative positions need to be manually defined by the user, and the results can vary significantly based on these decisions \cite{srinivasan2008statistical, Materka1998}. 

The final two methods are model based approaches and transform approaches. Model based methods utilize statistical models such as a Hidden Markov Model \cite{Bhat2016} to classify texture features from the gray level co-occurrence matrix. Lastly, the transform approach uses frequency methods such as Fourier or wavelet transforms to extract information about feature frequency or relative sizing in the texture \cite{Materka1998}. However, with transform methods, relative positioning of the texture features is lost in the process and further analysis is required to obtain this information \cite{Wang2008}. With all of the methods discussed so far, expert knowledge of the process/analysis is required for interpreting the results, and it is difficult to target a specific feature in a texture such as the specific pattern shape or depth of features. 

\begin{figure*}[htbp]
    \centering
    \includegraphics[width=\textwidth]{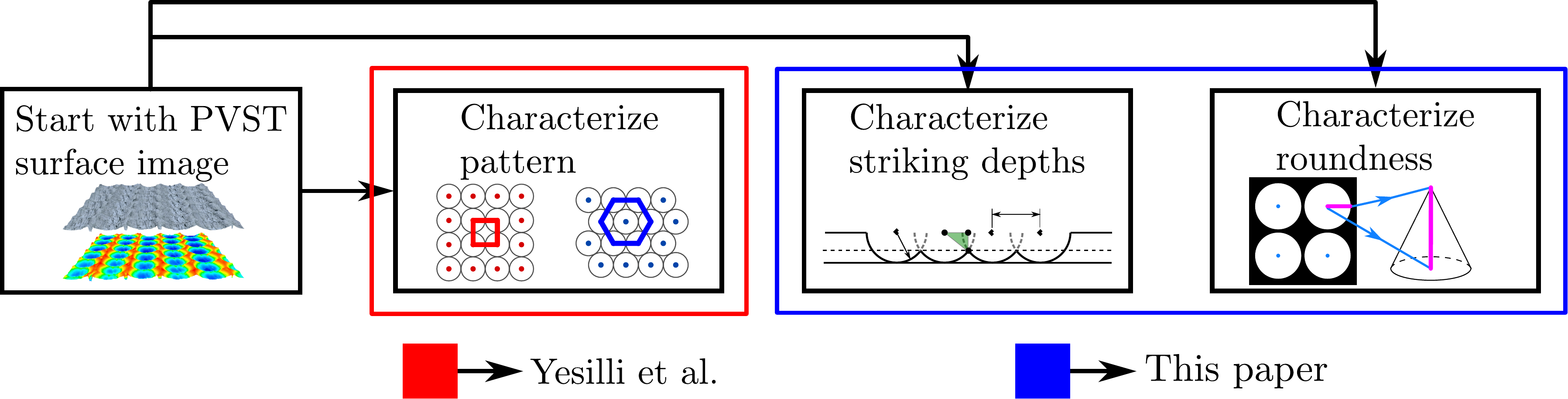}
    \caption{An overview flow chart for PVST texture characterization. Starting with a PVST image, three main features can be classified (depth, roundness, and pattern).}
    \label{fig:overview}
\end{figure*}

This paper describes a Topological data analysis (TDA) approach for quantifying surface texture and pattern, and it shows the validity of this approach by applying it to PVST surface images. 
Figure~\ref{fig:overview} shows an overview of the developed pipeline, and the first box in the figure shows an example surface image. 
While our prior work extended the the TDA approach in \cite{Motta2018} to classify surface patterns formed by the indentation centers in PVST processes \cite{Yesilli2022} (second box in Fig.~\ref{fig:overview}), quantifying the consistency of indentation depths (third box in Fig.~\ref{fig:overview}), and characterizing generalized radii of indentation shapes, e.g., the profile of the indenter at different heights (last box in Fig.~\ref{fig:overview}) are two important problems that have not been addressed before. 
Specifically, the striking \textit{depth} and \textit{roundness} of semi-spherical PVST indenters are essential for characterizing a PVST surface and they enable predicting the impact forces in the PVST process \cite{chen2021force}. Quantifying these properties allows process control and ensures consistent mechanical properties for the part, if the impact forces are constant from strike-to-strike. 
We provide a framework for automatically characterizing general patterned texture, and apply it to quantitatively describe PVST surfaces. Within this framework, we characterize striking depth and roundness from PVST surface images using sublevel persistent homology (a tool from TDA). Another contribution of this work is locating the specific feature depths to locate a reference height for the surface.
This enables not only quantifying the indentation roundness at different heights, but it also allows estimating surface deviations from the theoretical $z=0$ reference plane, e.g., the surface slope. 
The developed tools, along with our previously described method for quantifying the patterns of the indentation centers \cite{Yesilli2022}, provide a quantitative approach for characterizing surfaces from texture-producing processes such as PVST. 

We start by describing the PVST process in \ref{sec:pvst}. We then introduce the relevant TDA background followed by derivations for the theoretical expressions used for quantifying texture features in Section~\ref{sec:theory}. The results are presented in Section~\ref{sec:results}, and the concluding remarks are listed in Section~\ref{sec:conclusion}. Finally, CAD simulation of PVST patterns, a feature score noise analysis, and surface slope and angularity estimation are included in the appendices. 

\subsection{Piezo Vibration Striking Treatment (PVST)} \label{sec:pvst}
\begin{figure*}[htbp]
    \centering
    \includegraphics[width=0.9\textwidth]{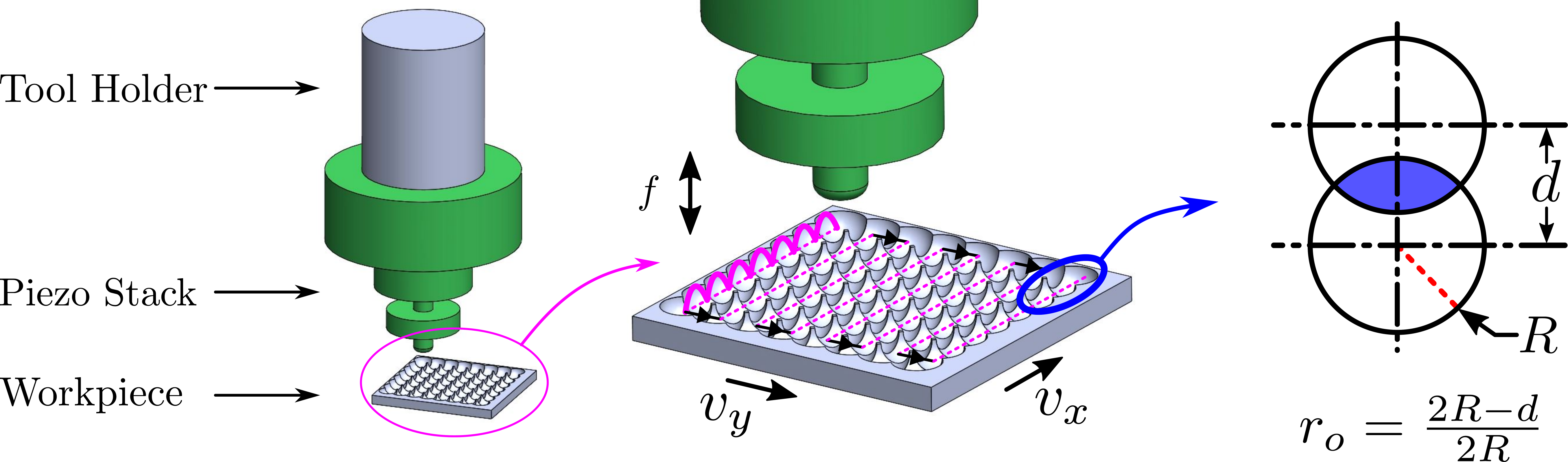}
    \caption{PVST diagram showing the mechanics of the PVST process and how the texture can be controlled using the frequency $f$, the in plane scanning speeds $v_x$ and $v_y$, and the overlap ratio $r_o$.}
    \label{fig:pvst_diagram}
\end{figure*}

 PVST is a process in which a piezo stack controlled with a CNC machine is used to impact the surface at a specific frequency leaving behind a surface texture on the part. Geometric characteristics of the texture are chosen by varying process parameters such as the shape of the indenter, the impact frequency, and scanning speeds. The diagram shown in Fig.~\ref{fig:pvst_diagram} demonstrates how the PVST process generates a texture on the surface as a result of the process parameters.
We see that the piezo stack produces oscillations in the impact tool that plastically deform the surface at regular intervals, and the stack is translated in the plane using the CNC machine to produce the texture. Parameters such as the oscillation frequency $f$, in-plane speeds $v_x$ and $v_y$, and the overlap ratio $r_o$ can be varied to produce different textures. As a result, it is important to be able to compare the output surface texture to the nominally expected texture based on the input process parameters. This comparison will allow for quantification of the process effectiveness and ensuring that the mechanical properties are within the expected tolerances compared to the results for the nominal texture.

\section{Background and Theory}
\label{sec:theory}
Section \ref{sec:PH} provides a brief background on persistent homology, the main tool from TDA that we use in this work. Sections \ref{sec:depth} and \ref{sec:roundness_expressions} show the derivations for the expressions that will be used to score the strike depths and roundness, respectively, of the PVST surfaces. Section \ref{sec:refHeights} shows how the process knowledge was applied to locate the strike minima and obtain the surface reference height. Section~\ref{sec:generalization} describes how our approach can be generalized to other tool shapes. 

\subsection{Persistent Homology Background} \label{sec:PH}
Persistent homology (PH) is a tool from topological data analysis (TDA) that allows for quantification of features in a data set by providing information about things like connectivity and loops in the data. We will describe PH through the lens of a PVST image rather than presenting abstract homology constructs, and we refer the reader to \cite{dey2022computational} for a comprehensive presentation of TDA. 

In this work, we use a specific type of PH called sublevel set persistent homology in which a height function is defined on the image. Let $I$ be the $p\times q$ image matrix of interest defined on the interval $[0,1]$. We define a parameter $T \in \mathbb{R}$ to be an arbitrary height in the image and $I_T = f^{-1}[0,T]$ to be a new image that is obtained by taking the sublevel set of $I$ up to a height $T$. Parameterizing the image sublevel sets allows for the topology to be studied as $T$ is varied using persistent homology. The topology is determined for each sublevel set of the image by only including pixels with gray scale values at or below the threshold $T$, and the homology is computed at each height \cite{Kaji2020}. This allows tracking the birth and death of \textbf{connected components} in the image, and the formation of \textbf{loops} in the process as $T$ is increased. 

\begin{figure*}[htbp]
\centering
\includegraphics[width=0.9\textwidth]{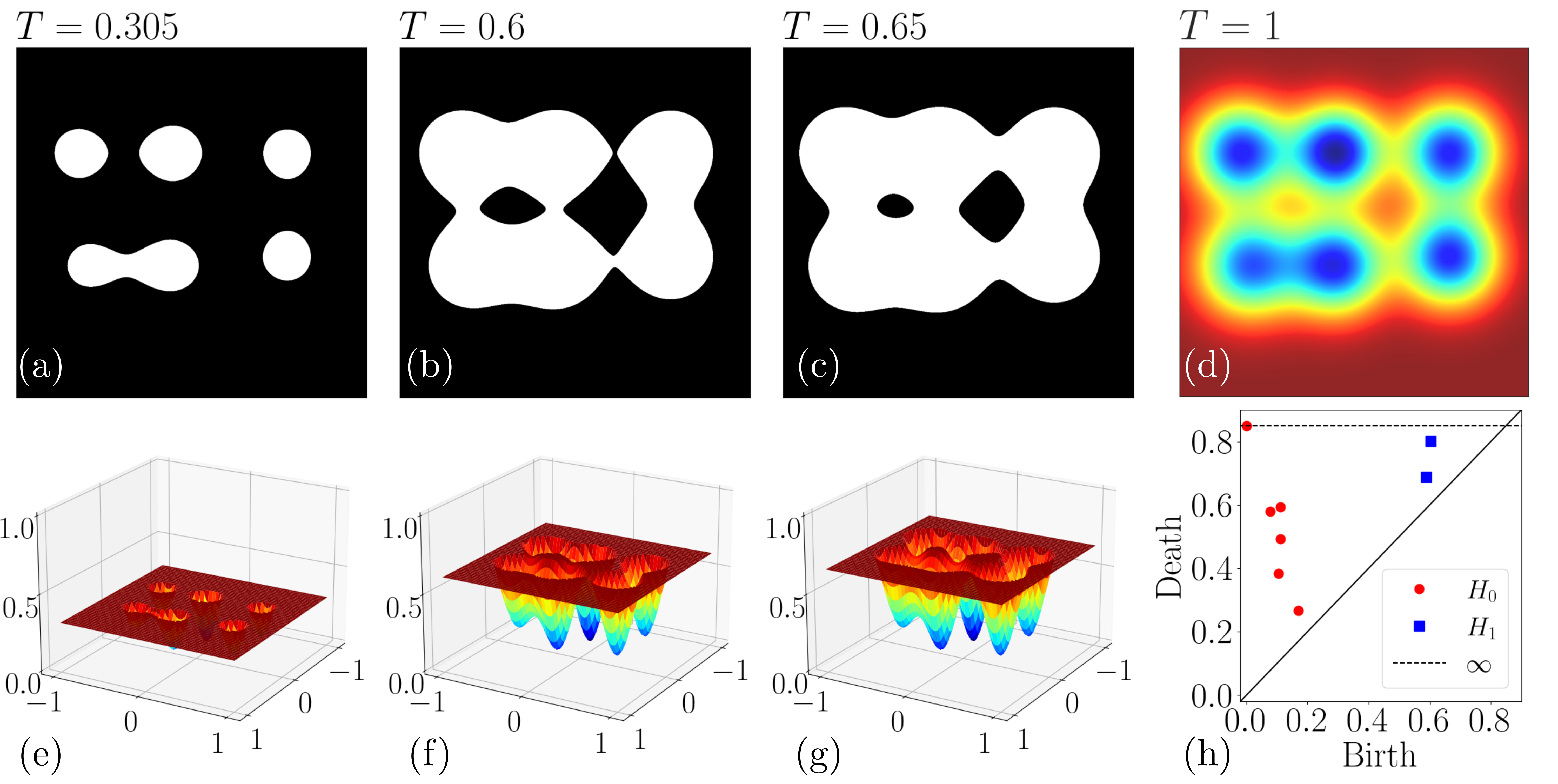}
\caption{Sublevel persistent homology example. (a-c) shows the surface level set represented as a binary image at 3 threshold heights, (d) shows the full surface image (e-g) shows the corresponding 3D surface plots for the thresholded images and (h) shows the full sublevel set persistence diagram for the surface.}
\label{fig:SL_pers_example}
\end{figure*}

We illustrate the concept of sublevel PH using a synthetic surface constructed by superimposing 6 Gaussian distributions as shown in Fig.~\ref{fig:SL_pers_example}(d). This surface can be compared to a PVST grid if each Gaussian distribution is imagined to be a strike in the PVST scan. The example image shows 6 prominent structures (blue) resulting from the Gaussian distributions and due to the relative positions, we see two loops in the image between the 6 components shown as orange circular structures. We will use sublevel persistent homology on this image to capture the aforementioned features in a quantifiable manner. The image in Fig.~\ref{fig:SL_pers_example}(d) was thresholded for all $T \in [0,1]$ and persistence was used to determine the image topology at each height and to track the formation of connected components and loops in the image. We note that the 0D homology or $H_0$ tracks the connectivity of the features and 1D homology or $H_1$ tracks the loops in the persistence diagram. The example in Figs.~\ref{fig:SL_pers_example}~(e-g) shows three different level sets for the full surface with corresponding binary images in Figs.~\ref{fig:SL_pers_example}~(a-c). Starting with Fig.~\ref{fig:SL_pers_example}(a), it is clear that 6 components were born in the image at this threshold, but two of them have connected or merged at this height. This connection is indicated in the persistence diagram by plotting the ($birth,~death$) coordinate for the younger of the two classes, i.e., the class that appeared at a higher $T$ value. We plot this connection as a red point with coordinate $(0.17, 0.27)$ in Fig.~\ref{fig:SL_pers_example}(h). Figure~\ref{fig:SL_pers_example}(b) shows that increasing $T$ to $0.6$ causes all 6 classes to connect into one component that persists to $\infty$. This is shown in the persistence diagram by plotting 4 more points at $(0.105, 0.38)$, $(0.11, 0.49)$, $(0.08, 0.56)$, and $(0.11, 0.59)$. The final red point indicates the infinite lifetime of the overall object on the dashed line. Note also that at a threshold of 0.58, the left loop is born meaning that a closed loop can be formed in the white region around a black region as shown in Fig.~\ref{fig:SL_pers_example}(b) at $T=0.6$. A second loop is born at 0.602 shown in Fig.~\ref{fig:SL_pers_example}(c) at $T=0.65$. When the threshold height reaches the point where the loops fill in with white in the level set, the loop dies and the point is plotted in the persistence diagram. For this example, the loops are born at 0.58 and 0.602, and die at 0.69 and 0.80 respectively as shown in Fig.~\ref{fig:SL_pers_example}(h). As $T$ reaches its highest value at 1 (Fig.~\ref{fig:SL_pers_example}(h)), the full persistence diagram is obtained. The loop on the right side is visually larger than the left one in Fig.~\ref{fig:SL_pers_example}(d), and this is indicated by the top blue square point having a larger distance to the diagonal in the persistence diagram giving that loop a longer \textit{lifetime}. The distribution of points in the persistence diagram can then be studied to compare to the expected distribution of persistence pairs for a nominal surface.

A major benefit of utilizing sublevel persistence to study various features of a function is that it has been shown to be stable under small perturbations due to noise \cite{CohenSteiner2006}. Specifically, the bottleneck distance between persistence diagrams is defined as $d_B(X,Y)=\inf_{\gamma} \sup_{x}||x - \gamma(x)||_\infty $ where $x\in X$ and $y\in Y$ are the persistence diagrams (birth and death coordinates) and $\gamma$ is the set of possible matchings between $X$ and $Y$. If one diagram contains more persistence pairs, those pairs are matched to the diagonal in $\gamma$. The main theorem in \cite{CohenSteiner2006} states that for two continuous well-behaved functions, $f$ and $g$, the bottleneck distance satisfies,

\begin{equation}\label{eq:pd_stability}
    d_B(D(f),D(g))\leq ||f-g||_\infty,
\end{equation}
where $D(f)$ and $D(g)$ are the sublevel persistence diagrams for $f$ and $g$. Assume that $f$ is the nominal texture surface and $g$ is the same texture that contains additive white noise. We represent the textures here as functions $f,g: \mathbb{R}^2\xrightarrow{} \mathbb{R}$ where the output of the functions is a depth map for the texture. Equation~\eqref{eq:pd_stability} states that the bottleneck distance between the nominal and noisy surface persistence diagrams will remain bounded by the largest deviation between the surfaces. This result allows for noise robust comparisons between the nominal and experimental texture persistence diagrams.

\subsection{Strike Depth}\label{sec:depth}

In order to compare the experimental persistence results with the nominal surface pattern, we need to derive expressions that describe the persistence of nominal patterns as a function of the process parameters. We start with the PVST strike depth, and we consider the scenario of deriving the sublevel persistence of a nominal PVST grid. 

\subsubsection{Theoretical Expressions}\label{sec:expressions_depth}
\begin{figure*}[htbp]
	\centering
	\includegraphics[width=0.9\textwidth]{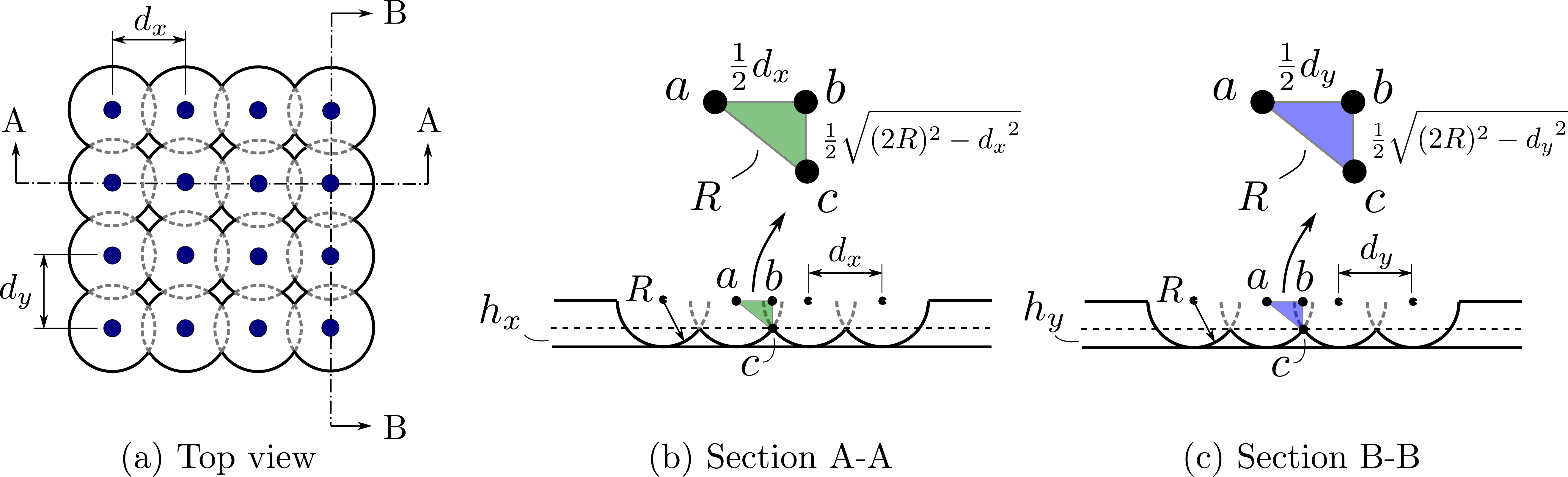}
	\caption{Arbitrary PVST lattice diagram with a grid top view (a), section views for the strike rows (b) and columns (c) to illustrate the geometry of a PVST grid,}
	\label{fig:sub_level_diagram}
\end{figure*}

Based on the PVST process inputs, we expect the ideal texture to consist of a square grid of overlapping circular indentations where all strikes have uniform depths. Consider the side views of a single row and column in a perfect PVST lattice with arbitrary overlap ratios in Fig.~\ref{fig:sub_level_diagram}.
\noindent
where $R$ is the nominal radius of the circle obtained from a PVST impact, $d_{x}$ and $d_{y}$ are the horizontal and vertical distances between centers accounting for overlap ratios. In general, the grid does not have to be square so we derive our expressions assuming a general grid shape and apply the special case for a square grid later. In the horizontal direction, the overlap ratio is defined by 
\begin{equation}\label{eq:overlap_ratiox}
		r_x=\frac{2R-d_{x}}{2R},
\end{equation}
where $r_x$ is the overlap ratio in the $x$ direction.
Using the geometric expressions in Fig.~\ref{fig:sub_level_diagram}, $h_x$ measured from the maximum depth of the impact can be computed using 
\begin{equation}\label{eq:height1}
		h_x=\frac{1}{2}\left( 2R-\sqrt{(2R)^2-{d_{x}}^2}\right),
\end{equation}
where $h_x$ is the height at which all of the impact \textbf{rows} merge. Combining Eq.~\eqref{eq:overlap_ratiox} and Eq.~\eqref{eq:height1} to eliminate $d_{x}$ gives an expression for the height $h_x$ in terms of the impact radius and the overlap ratio 
\begin{equation}\label{eq:height}
		h_x=R\left( 1-\sqrt{(2-r_x)r_x}\right).
\end{equation} 

Similar expressions can be obtained for the vertical direction by replacing $x$ with $y$. In order to normalize Eq.~\eqref{eq:height}, we rescale the radius of the PVST strikes at maximum depth to one. This is consistent with the PVST gray scale images used for the experimental analysis. This means that Eq.~\eqref{eq:height} can be normalized by setting $R=1$ as this makes the connecting height 1 for an overlap ratio of 0. The normalized heights will be denoted by $\overline{h_x}$ and $\overline{h_y}$, respectively, and can be computed using 
\begin{equation}\label{eq:normh}
	\overline{h}= 1-\sqrt{(2-r)r},
\end{equation}
where $\overline{h}$ is the height in the $x$ or $y$ direction as a function of the overlap ratio $r$ in the $x$ or $y$ direction respectively. Notice that Eq.~\eqref{eq:normh} achieves maximum value when $r$ is zero and minimum value when $r$ is one. 

\begin{figure}[ht]
	\centering
	\includegraphics[width=0.48\textwidth]{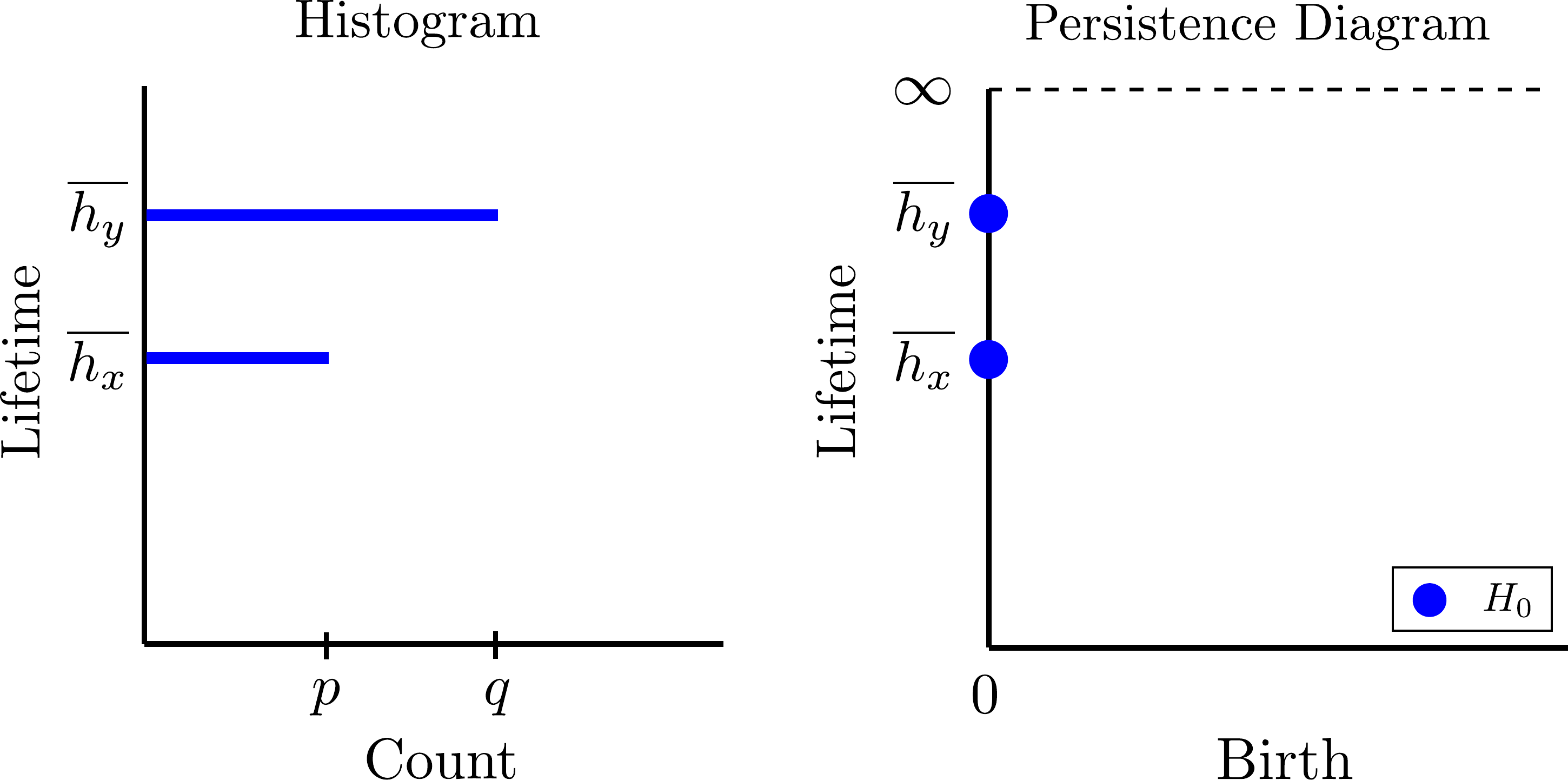}
	\caption{Theoretical sublevel persistence diagram and histogram for the striking depths for an arbitrary $p \times q$ grid with critical heights $h_x$ and $h_y$.}
	\label{fig:theoreticalpers}
\end{figure}

Without loss of generality, we assume that $r_x > r_y$. This means that the horizontal rows will connect before the columns because $h_x < h_y$. Therefore, if there are $p$ rows in the grid, $p$ classes will die at $h_x$ and if there are $q$ columns in the grid, $q$ more classes are expected to die at $h_y$ in the 0D persistence diagram when $p\times q$ classes are born at $h=0$. A theoretical persistence diagram was generated for the scenario when $q>p$ shown in Fig.~\ref{fig:theoreticalpers}, but in general the relative sizes of $p$ and $q$ can vary depending on the number of rows and columns in the grid.
For the images of interest, it was expected that the grid would be square ($p=q=n$) and the overlap ratio was constant in both directions ($r_x=r_y=r$). As a result, we expect $n^2$ classes to be born at 0 and die at a height $h$. Table~\ref{tab:expectedvalues} shows the expected lifetime of the PVST strikes for different overlap ratios using Eq.~\eqref{eq:normh}. See Section~\ref{sec:verification} for CAD-based simulations used to confirm the theoretical derivations. 

\begin{table}[htbp]
\caption{\label{tab:expectedvalues}Expected striking depth lifetimes for different overlap ratios where the grid is square ($n\times n$) strikes, and the heights have been normalized to correspond to a strike radius of 1.}
\centering
\begin{tabular}{|c|c|c|c|c|}
  \hline
  Overlap Ratio & 0\% & 25\%  & 50\% \\ 
  \hline
  Lifetime ($\overline{h}$) & $1$ & $0.339$ & $0.134$ \\ 
  \hline
\end{tabular}
\end{table}

\subsubsection{Depth Score}

In this section we develop a score to to quantify the uniformity of striking depths thus allowing a comparison between the experimentally measured depths and their nominal counterparts. We start by obtaining nominal and experimental histograms to show the sample distributions of the sublevel persistence lifetimes of the strikes. We plot probability density on the $x$ axis, and persistence lifetime on the $y$ axis where the experimental lifetimes come from a direct persistence computation on the image, and the nominal distribution is obtained from the theoretical expressions. Note that the number of histogram bins for the experimental images was determined using Rices Rule which states that the number of bins $k$ is computed using $k=\lceil 2\sqrt[3]{n}~\rceil$ where $n$ is the number of persistence pairs in the persistence diagram \cite{rices_rule}. Once the two distributions are obtained, we compute the Earth Movers Distance between them to quantify the differences between the distributions \cite{EMD}. A normalized score was desired to allow for comparison of the earth movers distances for the striking depth distributions. The Earth Movers Distance (EMD) can be analytically analytically computed according to
\begin{equation}\label{eq:emd}
	\mathrm{EMD}(u,v)=\inf_{\pi \in \Gamma (u,v)}\mathbb{E}_{(x,y)\sim \pi}[|x-y|],
\end{equation}
where $u$ and $v$ are the two distributions, $\mathrm{EMD}$ is the earth movers distance between $u$ and $v$, and $\Gamma$ is the set of distributions that exist between $u$ and $v$. In other words, the EMD computes the minimum amount of work required to transform one distribution into the other \cite{EMD}.

It should be noted that Eq.~\eqref{eq:emd} can be used to compare any two distributions $u$ and $v$, so it would be straight forward to directly compare the nominal and experimental persistence diagrams to measure the combination of feature depth and surface flatness. While this is a perfectly valid method for quantifying general differences in the textures, it does not directly provide information about a specific texture feature of interest as it considers both birth and death of the features. For this reason, it was chosen to consider the lifetime distributions as probability distributions to isolate the effect of the feature rather than where it is born in the image and provide a path for normalizing scores to quantifying these features in a way that is easy to understand for the user.

For the PVST striking depth distributions, the images have been normalized from 0 to 1. This means that each pixel can only contain a value in the finite interval 0 to 1. As a result, the maximum possible persistence lifetime for a feature occurs when the feature is born at zero and survives for the entire range of the height function. Conversely, the minimum persistence lifetime occurs when the feature is born at zero and survives for an infinitesimal time. The difference between these lifetimes corresponds to the maximal earth movers distance for any two images because Eq.~\eqref{eq:emd} is independent of the number of observations. This means that in this case, the earth movers distance has an upper bound where one distribution has all persistence pairs with lifetimes at 1 and the second distribution has all persistence pairs with 0 lifetime. Therefore, the maximum possible earth movers distance in this case is 1, and the distances computed for the different overlap ratios can be directly compared. We define the depth score $0 \leq \bar{D} \leq 1$ according to
\begin{equation}
	\bar{D}=1-\mathrm{EMD},
\end{equation}
where $\bar{D}=1$ when the actual depth distribution is identical to the expected distribution, while $\bar{D}=0$ when the distributions are the farthest apart. A score between 0 and 1 allows for characterizing the effectiveness of the PVST striking depth distribution as a percentage score where higher percentages indicate improved uniformity in the depth distribution of PVST strikes.

\subsection{Strike Roundness}\label{sec:roundness_expressions}

Since sublevel persistence does not encapsulate spatial information, it cannot be used by itself to characterize roundness of the PVST strikes. Therefore, we needed a tool that can encapsulate that information before using persistence to characterize the shape of the PVST strikes. The tool we used is the distance transform, which transforms each pixel of the image to display its euclidean distance to the nearest background pixel (black) as a gray scale intensity. Each image needed to be thresholded at a particular height to compute the distance transform, i.e., any pixel below the height is set to black (0) and any pixel above is set to white (1). The distance transform then sets each pixel to a gray scale value encoding that pixels minimum euclidean distance to the nearest black pixel. In other words, the image is transformed to show information about the size of the circles in the third dimension rather than the depths. To obtain theoretical results for quantifying the roundness of the strikes, we first needed to develop a transformation to convert a number of pixels into a physical distance as described by
\begin{equation}\label{eq:distToPix}
	 x = \frac{n_{p}w}{P},
\end{equation}
where $n_p$ is the number of pixels corresponding to distance $x$ in the image with $P\times P$ pixels and $w$ is the width or height of the image in any desired unit system. We note that $x$ and $w$ must have the same units. Using the nominal process parameters such as the in plane speeds, overlap ratio and frequency, the nominal circle radius can be computed. An example case for computing the nominal radius is as follows: For a frequency of $f$, a speed $v_x$ in mm/min, image width $w$ in mm, the nominal radius in mm can be computed using
\begin{equation}\label{eq:nom_rad}
	R = \frac{v_x}{120f}.
\end{equation}
  
The factor of $120$ is an artifact of the unit conversions from minutes to seconds and division by two to obtain the radius instead of the diameter.

The speed $v_x$ is dependent on the overlap ratio with the relationship
\begin{equation}
	v_x=3000(1-r),
\end{equation}
where 3000 mm/min is the speed that results in a 0\% overlap pattern at a frequency of $100$~Hz.

Substituting the frequency and speed expression into \eqref{eq:nom_rad}, we obtain an expression for the nominal circle radius in terms of the overlap ratio,
\begin{equation}
	R = \frac{1}{4}(1-r),
\end{equation}
where $r$ is the overlap ratio and $R$ is the nominal circle radius in mm at a PVST frequency of 100 Hz. We then threshold the texture at a height $T$ and compute the circle radius at the given height using the geometry shown in Fig.~\ref{fig:cone_no}(a). It is clear that as the image threshold height changes, the circle radius also varies due to the geometry of the strikes. 
Using the Pythagorean theorem we can obtain a relationship between $\sigma$, $h$ and $R$ as follows
\begin{equation}
	h^2+\sigma^2=R^2.
\end{equation}
Solving for $\sigma$ and setting $h=R-T$ yields the following expression for the nominal radius at a given threshold height:
\begin{equation}\label{eq:sigma_radius}
	\sigma = \sqrt{(2R-T)T},
\end{equation} 
where $T$ is the threshold height from the bottom of the strike in mm. Using this information, we can threshold the image at various heights and apply the distance transform to allow for sublevel persistence to be used for measuring the strike roundness. Basically, the distance transform is used to encode spatial information as height information, thus allowing us to leverage sublevel persistence for scoring strike roundness as described in the following sections.  

\subsubsection{Sublevel Persistence for no overlapping strikes ($T<\overline{h}$)}\label{sec:no_overlap}
\begin{figure*}[htbp]
\centering
\begin{minipage}[t]{0.8\textwidth}
\centering
\includegraphics[width=\textwidth]{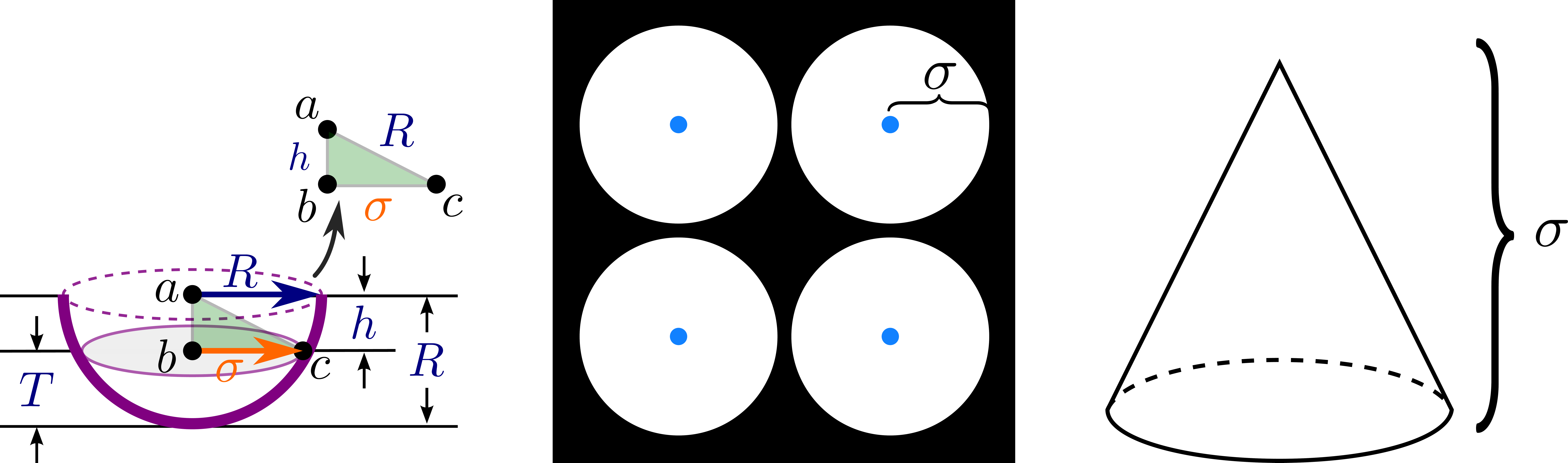}
\end{minipage}
\begin{minipage}{\textwidth}
\vspace{0.2 in}
\end{minipage}
\begin{minipage}[b]{0.26\textwidth}
\centering
(a) Impact Geometry
\end{minipage}
\begin{minipage}[b]{0.26\textwidth}
\centering
(b) Binarized Image
\end{minipage}
\begin{minipage}[b]{0.26\textwidth}
\centering
(c) Distance Transform
\end{minipage}
\caption{Converting between a binarized image and its corresponding distance transform geometry. (a) shows the strike geometry used for converting threshold heights to the radius of the strike at that height, (b) shows the thresholded image at a height below the critical height $T<\overline{h}$ (no overlap), (c) shows the cone geometry resulting from the distance transform of the binarized image and how the strike radius $\sigma$ appears in each form.}
\label{fig:cone_no}
\end{figure*}

Consider the PVST grid with no overlap shown in Fig.~\ref{fig:cone_no}(b). When the distance transform is applied to the thresholded grid, spatial information about the size of the circles is encoded as height information in the shape of cones (Fig.~\ref{fig:cone_no}(c)). As the distance from the edge of the circle increases, so does the height of the cones which can be understood from Fig.~\ref{fig:cone_no}(b). Applying sublevel set persistence to this grid of cones allows quantifying the roundness of the circles. 
We see that as the height of the connectivity parameter is varied starting at the bottom of the cones, 1 0D class is born at time 0 and remains to $\infty$. Applying one-dimensional persistence to the $n\times n$ grid of cones we expect $n^2$ 1D classes to be born at 0 and die at $\sigma$. 1D persistence was chosen for the roundness application because we are interested in the lifetimes of cycles in the images as they will provide information about the roundness of the strike. This was not necessary for the depth measurements because we only needed to know the depth at which the strikes connected.

\subsubsection{Sublevel Persistence for overlapping strikes ($T\geq \overline{h}$)}

\begin{figure*}[htbp]
\centering
\begin{minipage}[t]{0.8\textwidth}
\centering
\includegraphics[width=\textwidth]{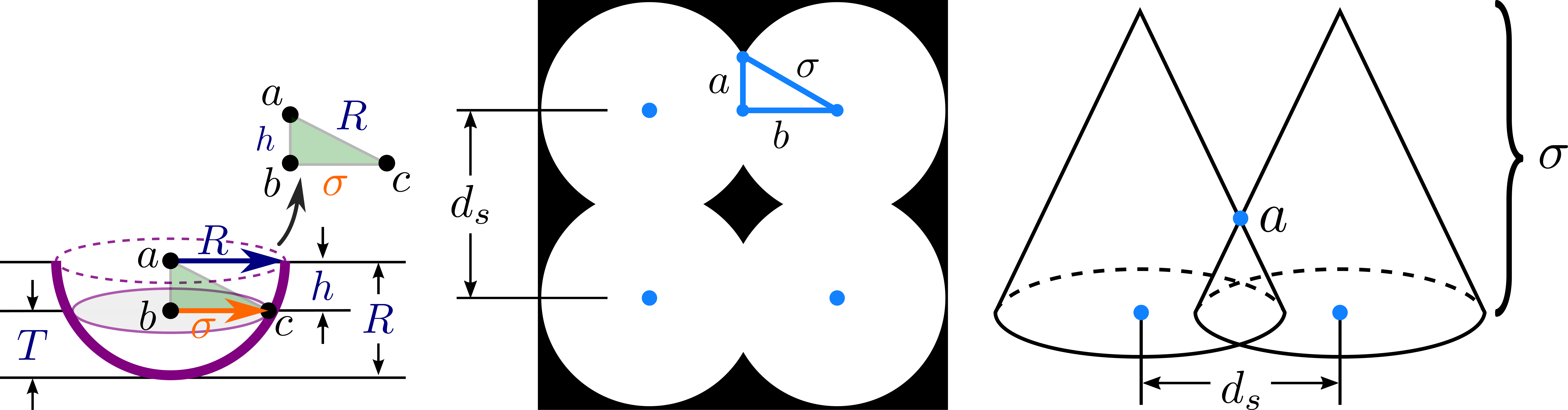}
\end{minipage}
\begin{minipage}{\textwidth}
\vspace{0.4 in}
\end{minipage}
\begin{minipage}[b]{0.3\textwidth}
\centering
(a) Impact Geometry
\end{minipage}
\begin{minipage}[b]{0.3\textwidth}
\centering
(b) Binarized Image
\end{minipage}
\begin{minipage}[b]{0.3\textwidth}
\centering
(c) Distance Transform
\end{minipage}
\caption{Converting between a binarized image and its corresponding distance transform geometry. (a) shows the thresholded image at a height below the critical height $T\geq \overline{h}$ \textbf{(overlap present)}, (b) shows the cone geometry resulting from the distance transform of the binarized image and how the strike radius $\sigma$ appears in each form, and (c) shows the geometry used to obtain expressions for the cone intersection height $a$.}
\label{fig:cone_overlap}
\end{figure*}

We now generalize the result from the case with no overlap by thresholding the image above the critical height ($\bar{h}$) where we obtain an image with overlapping circles shown in Fig.~\ref{fig:cone_overlap}(b). The critical height is the depth at which water would overflow from the strike into the other strikes and it can be computed using,
\begin{equation}
	\bar{h} = R(1-\sqrt{(2-r)r)},
\end{equation}
where $\overline{h}$ is the critical height, $R$ is the nominal circle radius, and $r$ is the overlap ratio. We define a new parameter $\epsilon$ to indicate the threshold height $T$ in terms of the critical height $\bar{h}$ using $T=\epsilon \bar{h}$ where $\epsilon$ defines the threshold height relative to the critical height. If $\epsilon < 1$, the circles in the thresholded image do not overlap and the case from Section~\ref{sec:no_overlap} is used, whereas if $\epsilon \geq 1$, the circles will overlap and a more general relationship needs to be considered. In Fig.~\ref{fig:cone_overlap}(a) we show a binarized image where $\epsilon > 1$ with strike overlap. When this thresholded image is distance transformed, a result similar to Fig.~\ref{fig:cone_overlap}(c) is obtained where a critical distance $a$ needs to be considered. The distance $a$ is the height in the distance transformed image where the gap between the cones connects to the surrounding object. Above $a$, the circles also disconnect in the filtration so we have a formation of cycles that can be considered when performing sublevel persistence.
To obtain an expression for $a$, we consider the triangle shown in Fig.~\ref{fig:cone_overlap}(b) and apply the Pythagorean theorem
\begin{equation}\label{eq:a1}
	a=\sqrt{\sigma^2-b^2}.
\end{equation}
An expression was needed for the side length $b$ in terms of other known parameters. For this, the center-to-center distance $d_s$ of the circles was used because we know that $d_s=2R(1-r)$ from the definition of the overlap ratio. At this point it is important to note that this expression depends on the full nominal radius $R$ and should not be written in terms of $\sigma$ because $d_s$ remains invariant for all threshold heights. Observe that, $d_s=2b$ from Fig.~\ref{fig:cone_overlap}(b) due to the circle position remaining constant. Applying the definition of $d_s$ to the result for $b$ we obtain an expression for $b$ in terms of known parameters
\begin{equation}
	b = R(1-r).
\end{equation}
Substituting $b$ into Eq.~\eqref{eq:a1} gives the critical distance
\begin{equation}
	a = \sqrt{\sigma^2-R^2(1-r)^2}.
\end{equation}

\subsubsection{Effect of High Threshold}

\begin{figure*}[htbp]
	\centering
	\includegraphics[width=0.6\textwidth]{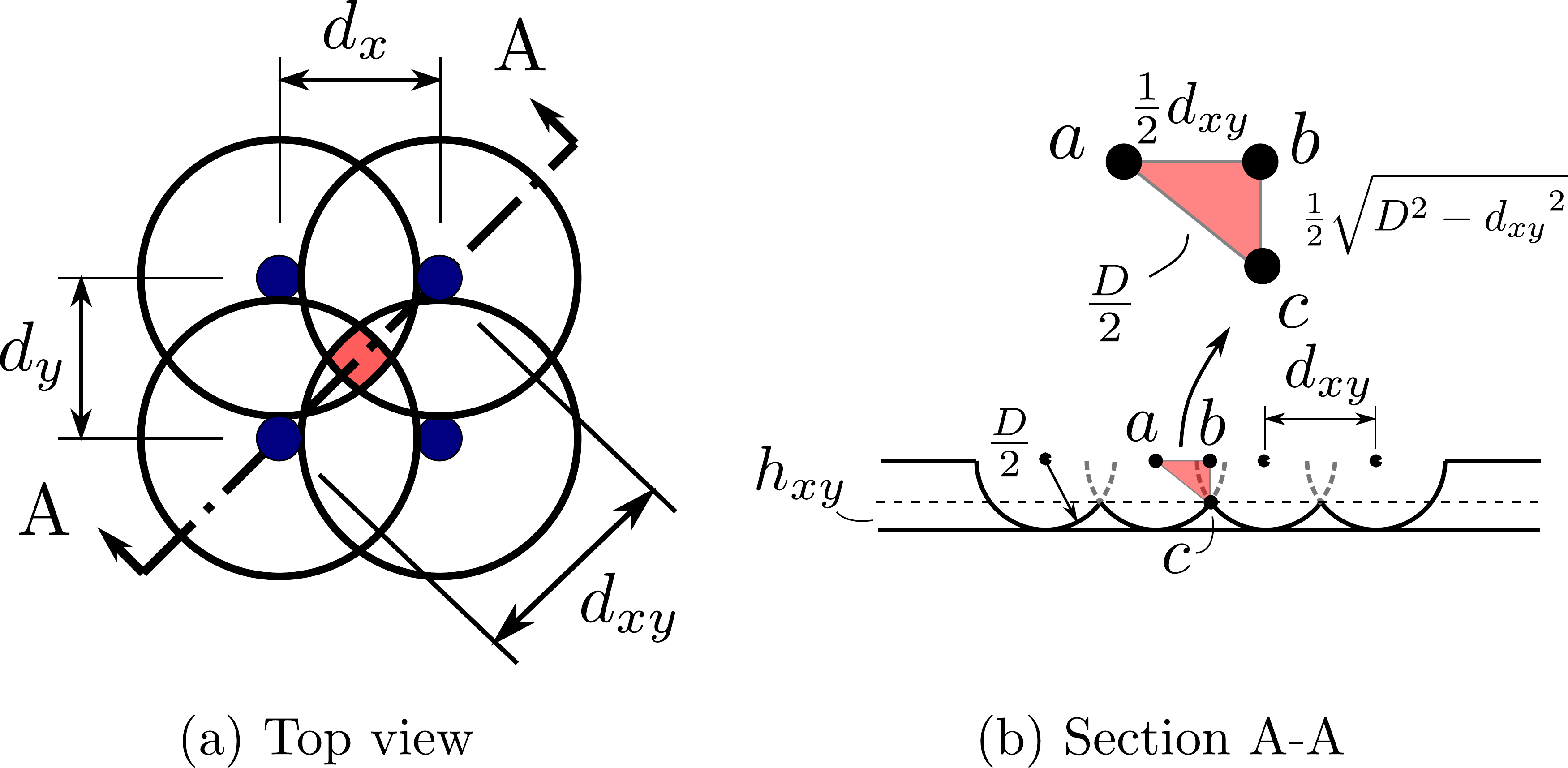}
	\caption{Nominal PVST grid with a high overlap ratio to demonstrate diagonal cross section overlap height.}
	\label{fig:hxy}
\end{figure*}

Lastly, we consider the gaps between the cones at higher overlap ratios. For low overlap ratio, the gap heights will span the entire depth of the strike, but as the overlap ratio increases, the gap height eventually begins to decrease causing the cycles to have lower lifetimes. To quantify this result, we needed to compute the height of the gaps as a function of the overlap ratio. Consider the grid diagonal cross section shown in Fig.~\ref{fig:hxy}.
We see that this section view results in the same triangle that was used to determine the closing height when categorizing the striking depths with the difference being the addition of the value $d_{xy}$. This value can be computed using
\begin{equation}
	d_{xy} = \sqrt{d_x^2+d_y^2},
\end{equation}
or if the grid is square
\begin{equation}\label{eq:dxy}
	d_{xy}=2\sqrt{2}R(1-r),
\end{equation}
where $R$ is the nominal strike radius. If we apply the Pythagorean theorem in the same way as the depth results, we obtain an expression for $h_{xy}$,
\begin{equation}
	h_{xy}=R(1-\sqrt{1-2(1-r)^2}).
\end{equation}
Substituting for an overlap ratio of 0.5, and the nominal radius of 1 due to the normalized depths, we obtain a value of $h_{xy}=0.29289$. It should be noted that $h_{xy}$ will be equal to the nominal radius $R$ as long as the following inequality is satisfied,
\begin{equation}
	d_{xy} > 2R.
\end{equation}
Here, if we assume equality, and substitute Eq.~\eqref{eq:dxy}, we find that this corresponds to an overlap ratio of $r=0.29289$. Because the 50\% overlap ratio case is larger than 29.28\% overlap, we needed to consider that all of the 1D persistence loops merge into a single loop above the height $h_{xy}$ whereas this phenomena was not present in the lower overlap ratio cases. This single component will have zero lifetime if the grid continues on forever.

\subsubsection{Roundness Expected Results Summary}
We summarize the PVST roundness expected 1D persistence results for the distance transformed images as follows:
\begin{enumerate}
	\item If $\epsilon < 1$, we expect $n^2$ classes to be born at $0$ and die at $\sigma$.
	\item For $\epsilon > 1$, we expect $1$ class to be born at $0$ and die at $\sigma$, and $n^2-1$ classes to be born at $a$ and die at $\sigma$.
	\item If $r > \frac{2 - \sqrt{2}}{2} \approx 0.29$ and $T>h_{xy}$, we expect one object to be born at time 0 and die at 0.
\end{enumerate}
See Appendix~\ref{sec:verification} for CAD-based simulations that confirm the theoretical results.

\subsubsection{Finding Reference Heights}\label{sec:refHeights}

\begin{figure*}[htbp]
	\centering
	\begin{minipage}{\textwidth}
	\centering
	\includegraphics[width=0.5\textwidth]{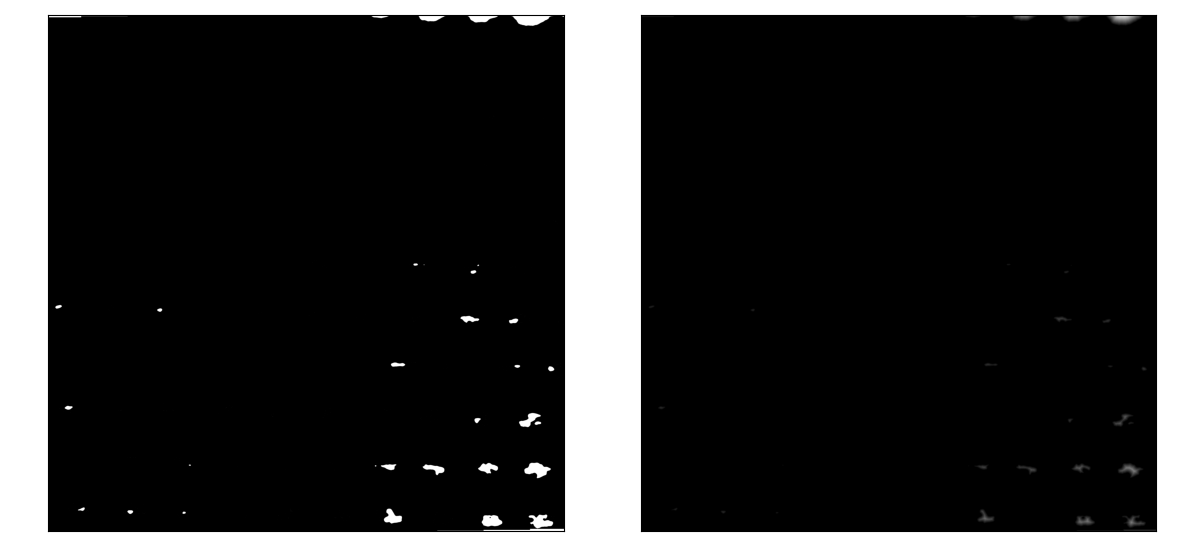}
	\end{minipage}
	\begin{minipage}{0.25\textwidth}
	\centering
	(a) Binarized Image
	\end{minipage}
	\begin{minipage}{0.25\textwidth}
	\centering
	(b) Distance Transform
	\end{minipage}
	
	\caption{50\% overlap ratio image thresholded at the reference height ($T=0.24$) found by taking the first threshold height that contained $19^2=361$ features in the persistence diagram. The binarized image is shown on the left and distance transformed image on the right.}
	\label{fig:ref_thresh50}
\end{figure*}

Due to variations in strike forces, initial surface heights, and artifacts in the images the strike minima do not lie uniformly at a height of 0 in practice, so a reference plane is required to determine what height to compare the roundness results with using the theoretical model. If the reference height is not used, then a shift would be present in the results that would skew the final roundness measurements. The first attempt at locating a reference height was to use the first height at which the persistence diagram contained a number of pairs equal to the number of strikes in the image. The problem with this approach is that the features that were obtained were due to noise in the image and very few of the strike minima were present in the image as shown in Fig.~\ref{fig:ref_thresh50} where we see the first threshold height in the 50\% overlap image that contains $19^2=361$ features. It appears that no strikes have been located in the top left corner of the image so this would be a poor estimate of the reference height for this image. 

In order to obtain a better reference height, we needed to first utilize knowledge of the surface to filter the persistence diagram down to the features that corresponded to the strike minima of the surface.

\begin{figure*}[t]
\centering
\begin{minipage}[t]{\textwidth}
\centering
\includegraphics[width=0.9\textwidth]{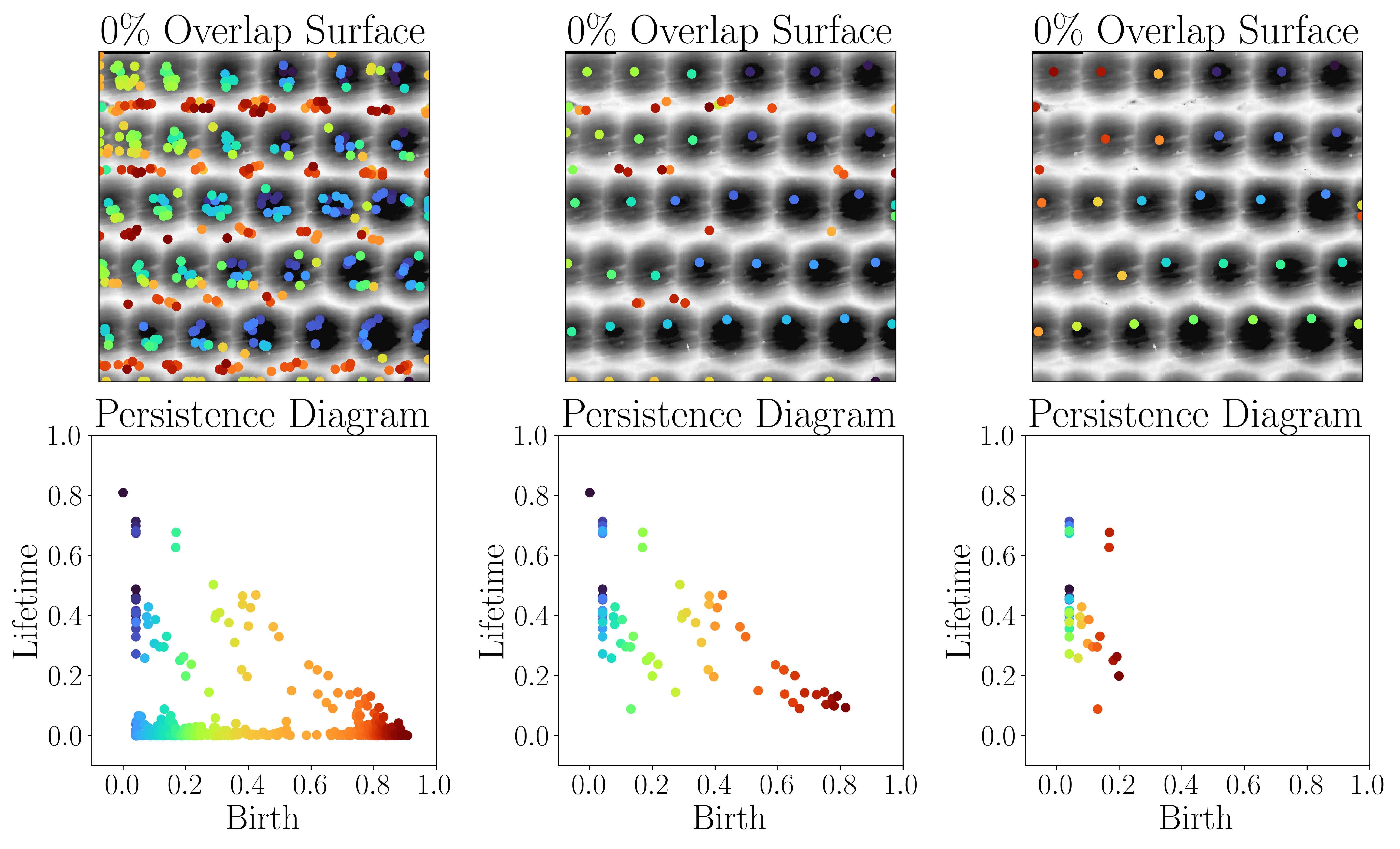}
\end{minipage}
\begin{minipage}{0.3\textwidth}
\centering
(a) Original Simplified Surface Persistence Diagram.
\end{minipage}
\begin{minipage}{0.3\textwidth}
\centering
(b) Low Lifetime Filtered Persistence Diagram.
\end{minipage}
\begin{minipage}{0.3\textwidth}
\centering
(c) Final Filtered Persistence Diagram.
\end{minipage}

\caption{Persistence diagram (PD) filtering on the simplified surface to locate strike minima. (a) The original PD of the simplified surface, (b) shows the persistence features after removing low lifetime features, and (c) shows the final filtered PD.}
\label{fig:pd_filtering}
\end{figure*}

\begin{figure*}[t]
\centering
\begin{minipage}[t]{\textwidth}
\centering
\includegraphics[width=0.48\textwidth]{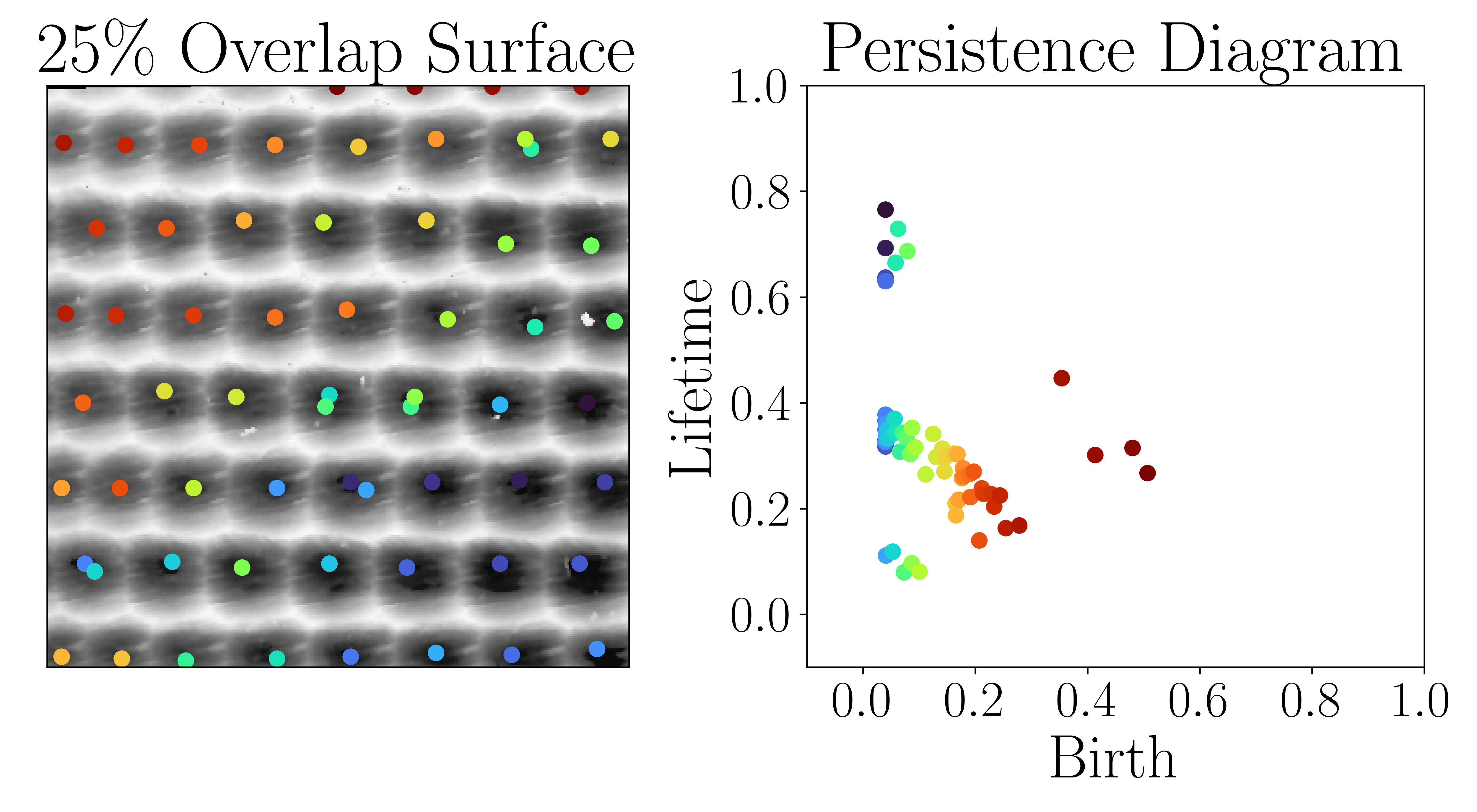}
\includegraphics[width=0.48\textwidth]{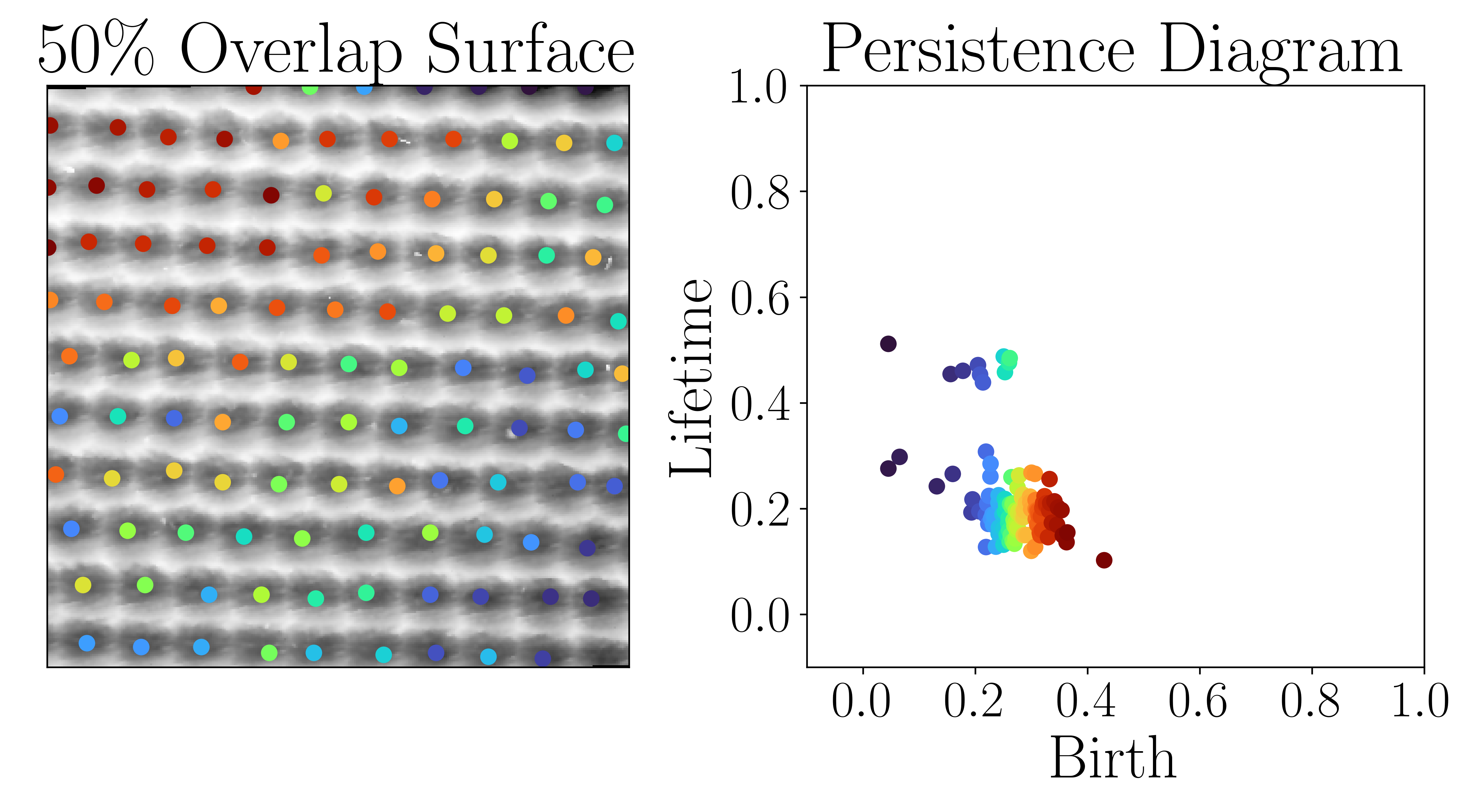}
\end{minipage}

\begin{minipage}{0.48\textwidth}
\centering
(a) 25\% Overlap Filtered Persistence Diagram.
\end{minipage}
\begin{minipage}{0.48\textwidth}
\centering
(b) 50\% Overlap Filtered Persistence Diagram.
\end{minipage}

\caption{Persistence diagram (PD) filtering on the simplified surface to locate strike minima. (a) 25\% Overlap filtered PD, (b) 50\% overlap filtered PD.}
\label{fig:other_filtered_pds}
\end{figure*}

We locate the strike minima by computing sublevel persistence on the surface and taking the birth times to be the minima of each feature. These points are plotted as shown in Fig.~\ref{fig:pd_filtering}(a). The critical points have been matched to their location in the persistence diagrams by color. From the color coding, it was clear that the blue/purple/green features corresponded to the strikes and the orange/yellow/red features corresponded to locations between the strikes. This observation allowed for the persistence diagram to be filtered in order to obtain the features of interest. Note that these images have been down sampled to 300$\times$300 down from 6000$\times$6000 to reduce the number of features in the image. The process begins by observing that there were approximately 35 features in this image, so the goal was to algorithmically filter the persistence diagram such that the resulting 35 features correspond to the actual strike minima. We start by filtering out low lifetime persistence pairs by computing a histogram of the lifetimes, and thresholding the lifetime above any point that contained a bar height larger than the number of desired features. This threshold resulted in the persistence diagram shown in Fig.~\ref{fig:pd_filtering}(b). It is clear that the features removed up to this point are attributed to noise as we see that each strike still retained at least one critical point after this step. We also observe that the features born at exactly time zero are due to the artifacts in the image, so the birth times were restricted to be larger than 0. The final step is to remove critical points from the right of the persistence diagram (red region) until only the desired number of features remain in the image; the result of this step is shown in Fig.~\ref{fig:pd_filtering}(c). The remaining features in the final filtered persistence diagram are taken to be the strike minima and the average height of these points is used as the reference height. Applying this process to the 25\% and 50\% overlap images yielded the results in Fig.~\ref{fig:other_filtered_pds}. We see that the located features in the filtered persistence diagrams are exceedingly close to the true strike minima and taking the average height of these points provided a good estimate of the reference zero height. A byproduct of this process is to enable estimating the surface slope/angularity by computing a regression plane to using the strike minima, as shown in Appendix~\ref{sec:slopes}.

\subsubsection{Roundness Score}
To quantify the feature roundness, the images needed to be thresholded at many different heights to compare the shapes to the nominal distribution over the entire feature. The output of this process is a curve for the earth movers distance as a function of the threshold height of the image. The overall feature roundness is then summarized by computing the area under this curve and diving by the interval width to remove the effects of different reference heights. For a general impact geometry, the area under this curve can be computed using Eq.~\eqref{eq:gen_round_score},

\begin{equation}\label{eq:gen_round_score}
	\bar{R}_G = \frac{1}{1-h_r}\int_0^1 \mathrm{EMD}(T)~\mathrm{d}T,
\end{equation}
where $h_r$ is the reference height of the image and $\bar{R}_G$ is the generalized roundness score for the texture. We note that this score, by definition, results in a larger score meaning that the texture shape is further from nominal and a lower score is closer to nominal. 

In order to obtain a roundness relationship similar to the percentage based depth score, we need to define a roundness score that is specific to the spherical impact by normalizing the area with an upper bound earth movers distance. Similar to the depth score, the earth movers distance at any threshold height is bounded above by two images with all pixels differing by the maximal distance between gray scale intensities. However, the distance transform operation makes it difficult to determine the maximum possible difference in pixel intensities because it is not possible to have all distances at the same value if at least one background pixel exists in the image. 

\begin{figure}
    \centering
    \includegraphics[width=0.3\textwidth]{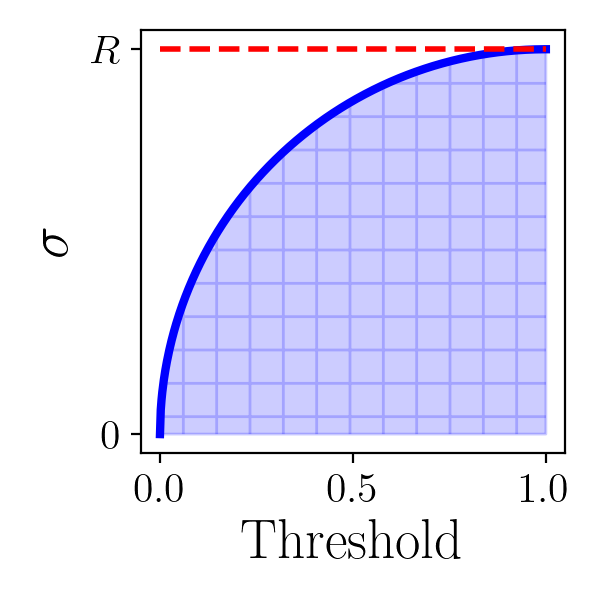}
    \caption{Plot of $\sigma$ as a function of the image threshold height to demonstrate the worst case earth movers distance plot.}
    \label{fig:worst_case_emd}
\end{figure}

To mitigate this issue, we assume for a reasonably generated physical texture, that the features will be generally close in size to the nominal features. To quantify this assumption, we will say that the experimental feature sizes will have a radius that is at most one nominal radius larger or smaller than the nominal feature size. By assuming that the experimental features are reasonably close in size to the nominal texture, it allows for the earth movers distance to be bounded by the radius at each threshold height and permits the definition of a percentage based score for this feature. For each threshold height of the image, the nominal radius is defined by $\sigma$. The $\sigma$ curve for a spherical feature geometry is defined by Eq.~\eqref{eq:sigma_radius} as a function of the height $T$ ranging from 0 to $R$ where $R$ is the maximum strike radius. We then introduce the change of variables $T=Rt$ where $t\in[0,1]$ is the image threshold height. This change of variables results in a quarter elliptical curve describing the maximum earth movers distance as a function of threshold height shown in Fig.~\ref{fig:worst_case_emd}.

The area under this quarter ellipse is computed as $\frac{\pi}{4}R$. A roundness score is then defined by normalizing the area under the experimental EMD curve by the quarter ellipse area and subtracting the result from unity to provide a percentage based score similar to the depth score. Equation \eqref{eq:sphere_rd_score} shows the spherical impact roundness score as a percentage where a higher score corresponds to the feature roundness being closer to nominal.
\begin{equation}\label{eq:sphere_rd_score}
    \bar{R}=1-\frac{4\bar{R}_G}{\pi R}
\end{equation}
The resulting score is specific to the spherical impact shape, and if a score is desired for a different impact shape, the $\sigma$ curve specific to that geometry needs to be obtained that bounds the earth movers distance and the score can be computed in a similar fashion. Note also that if the input experimental texture contains features that differ significantly from nominal this score will be less than zero so it should only be used on textures with feature sizes close in size to nominal. However, the generalized roundness score $\bar{R}_G$ can be used for any such texture, but a lower score means that the texture is closer to nominal in this case. See Appendix~\ref{sec:noise} for a quantification of the noise robustness of the depth and roundness scores.

\subsection{Generalizing for Other Textures} \label{sec:generalization}
\begin{figure}[htbp]
    \centering
    \begin{minipage}[t]{0.2\textwidth}
        \centering
        \includegraphics[width=\textwidth]{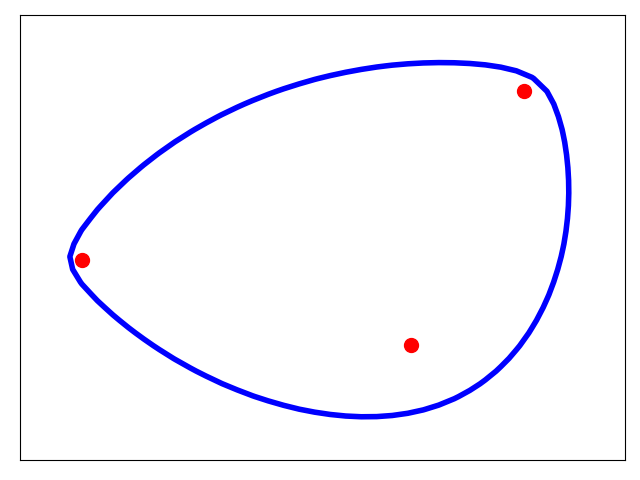}
    \end{minipage}
    \begin{minipage}[t]{0.2\textwidth}
        \centering
        \includegraphics[width=\textwidth]{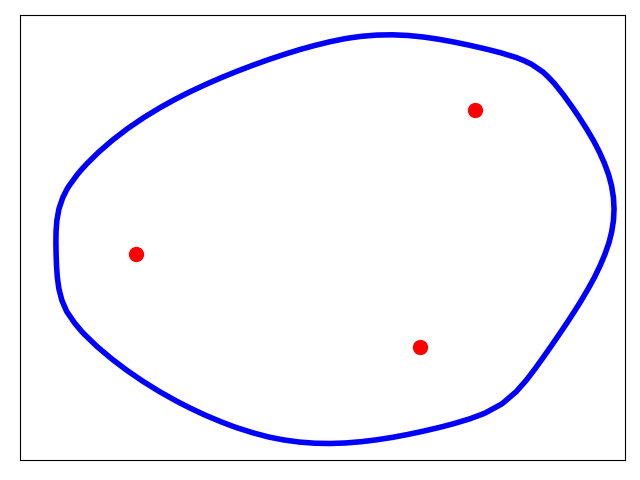}
    \end{minipage}
    \begin{minipage}[t]{0.2\textwidth}
        \centering
        \includegraphics[width=\textwidth]{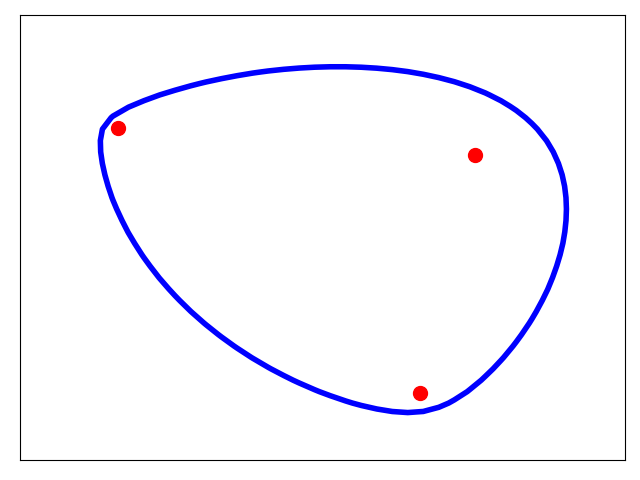}
    \end{minipage}
    \begin{minipage}[t]{0.2\textwidth}
        \centering
        \includegraphics[width=\textwidth]{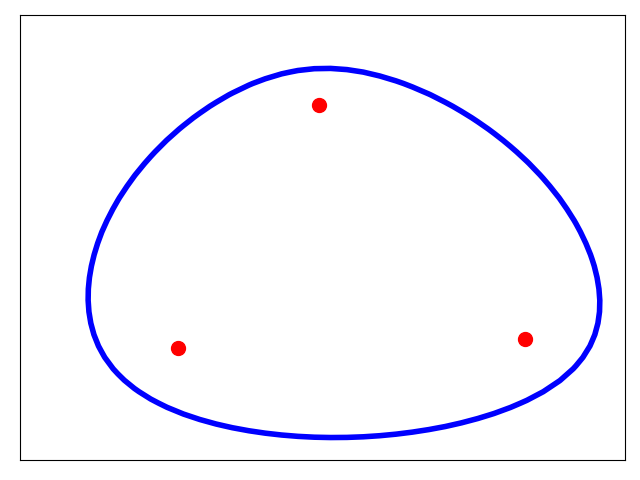}
    \end{minipage}
    
    \caption{Example plots of generalized conic sections to demonstrate different tool shape configurations for generalizing the results in this paper. Red points are the focal points of the conic, and the blue curve represents the cross section of the impact tool.}
    \label{fig:gen_conics}
\end{figure}

While the methods used in this paper were designed to account for features in a PVST image created using a semi-spherical tool, the process can be modified to account for any tool shape. One such example of a generalization of this process arises when a 5-axis milling machine is used to generate a dimple texture on a part. This process leaves behind elliptical dimples which result in improved texture properties \cite{Arizmendi2019}. It is clear that the methods used for analyzing a PVST texture will not work for this case. Generalizing the expressions used may introduce significant complexities in the analysis, but we provide two potential avenues for doing this. The first method offered is to apply the techniques in Appendix~\ref{sec:verification} where a CAD model is created for the nominal texture and the nominal persistence diagrams can be computed directly from the images for comparison with experimental results. This method is the most straight forward and has been shown to provide results within $5\%$ of the true values for the examples considered in this paper. The second method is to derive expressions for the theoretical persistence lifetimes using a generalized conic section to define the cross section shape. Pattern and depth can apply to any texture being analyzed, but roundness may not be a valid descriptor of the impact shape if it is not spherical. We adopt a \textit{generalized radius} feature that applies to a significantly larger set of indenter geometries to be the generalized conic section \cite{Grob1998} described by 

\begin{equation}\label{eq:gen_rad}
    \rho (x,y)=\sum_{i=1}^n\alpha_i||\vec{x}-\vec{b}_i||_p,
\end{equation}
where $\rho(x,y)$ is the generalized radius as a function of $x$ and $y$, $\alpha_i$ is the $i$th weight coefficient, $\vec{x}$ is a vector of coordinates ($(x,y)$ in this case), $\vec{b}_i$ is the $i$th focal point of the curve, and $p$ is the corresponding p-norm of the vector. For the special case of $n=1$, $\alpha=1$, $p=2$, and $b$ is the center point, we get the equation of a cone which has cross-sections of circles at various heights. Varying the weights and adding more focal points allows for arbitrary shapes to be formed such as the curves shown in Fig.~\ref{fig:gen_conics}.

\section{Results}
\label{sec:results}

The theoretical approaches were implemented on three PVST scans at varying overlap ratios being 0\%, 25\% and 50\% to quantify the strike depth and roundness in comparison to the respective nominal textures. We begin by measuring the strike depths for each image. 

\subsection{Strike Depth Results}
\begin{figure*}[htbp]
\centering
\includegraphics[width=0.8\textwidth]{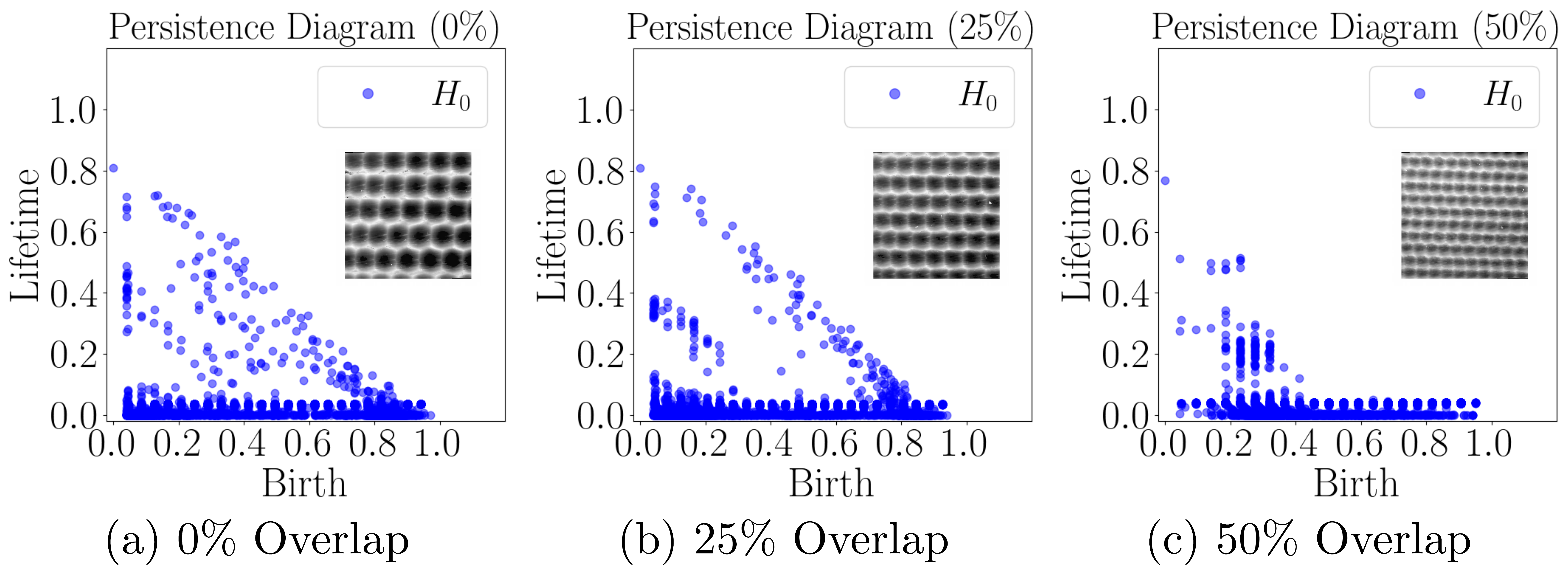} 
\caption{Unaltered striking \textbf{depth} persistence diagrams (a) 0\% Overlap PVST Image, (b) 25\% Overlap PVST Image, (c) 50\% Overlap PVST Image}
\label{fig:depth_pds}
\end{figure*}

Sublevel set persistence was applied to the PVST images shown in Fig.~\ref{fig:depth_pds} with the corresponding persistence diagrams adjacent to each image. We see a significant portion of the persistence pairs have negligible lifetime and are likely a result of noise in the images. The noise was removed from these persistence diagrams by generating histograms for the pairs and increasing the persistence lifetime threshold if any of the histogram bars had a count larger than the number of strikes in the image. This method is reliant on the observation that a large number of points are present in the low lifetime region of the persistence diagrams. We also filter by the birth times of the features by removing features with the largest birth times until the desired number remain similarly to Fig.~\ref{fig:pd_filtering}.

\begin{figure*}[t]
\centering
\includegraphics[width=0.8\textwidth]{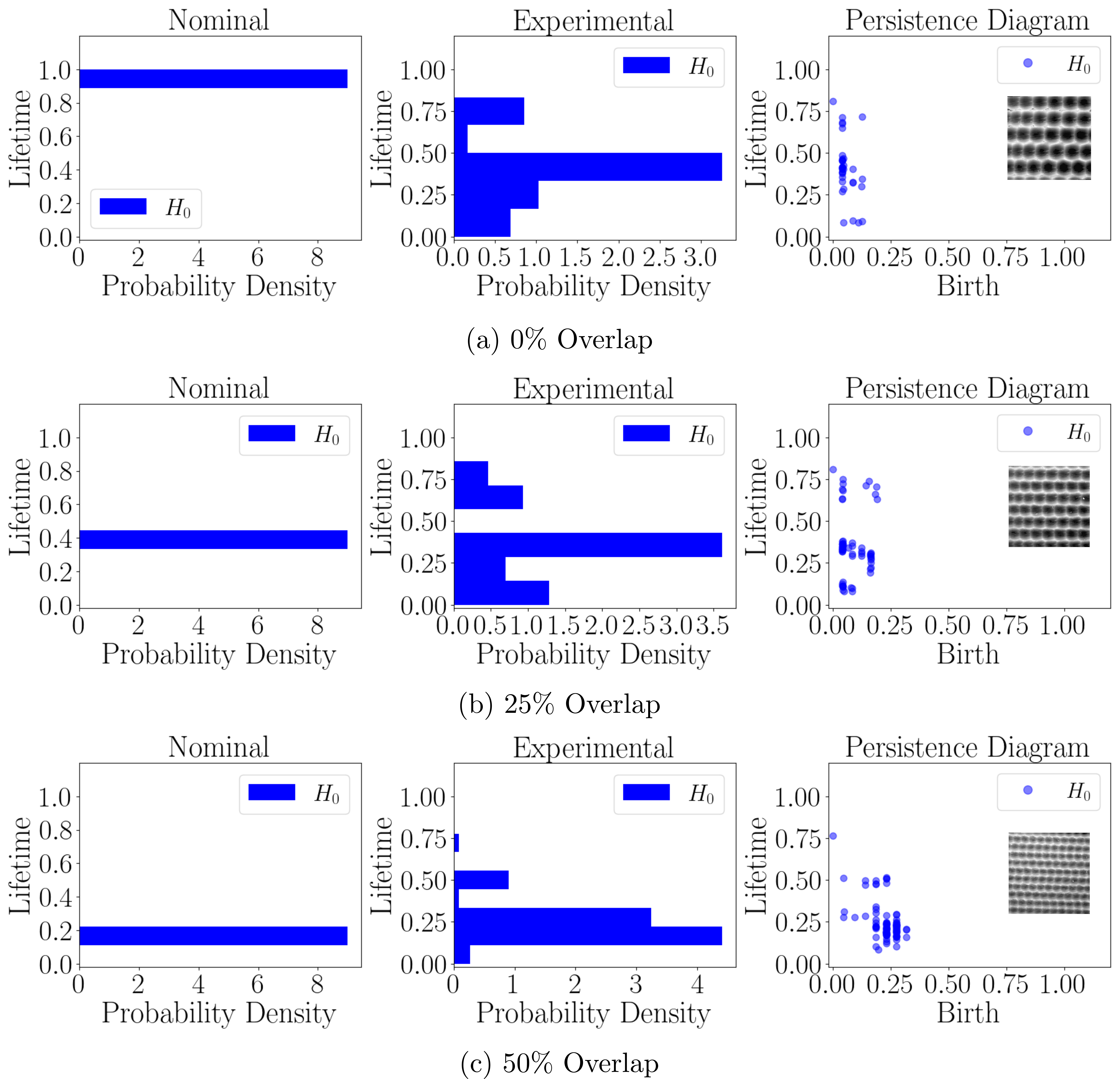} 
\caption{Noise filtered \textbf{depth} persistence diagrams and histograms for each overlap ratio.}
\label{fig:filtered_depths}
\end{figure*}

Applying these operations to the diagrams in Fig.~\ref{fig:depth_pds}, the filtered persistence diagrams in Fig.~\ref{fig:filtered_depths} were obtained. The corresponding computed depth scores are shown in Table~\ref{tab:depth_scores}, and it was clear that the 25\% and 50\% overlap images had significantly higher depth scores which could be a result of the strikes being closer together. 


\begin{table}[t]
\caption{\label{tab:depth_scores}Striking depth scores for each overlap ratio. Higher is closer to nominal.}
\centering
\begin{TAB}(r,0.1cm,0.1cm)[5pt]{|c|c|c|c|}{|c|c|}%
  Overlap Ratio & 0\% & 25\%  & 50\%\ \\ 
   $\bar{D}$ & $41.04\%$ & $86.31\%$ & $88.63\%$ \\
\end{TAB}
\end{table}

\subsection{Strike Roundness Results}

\begin{figure*}[htbp]
\centering
\begin{minipage}[t]{\textwidth}
\centering
\includegraphics[width=\textwidth]{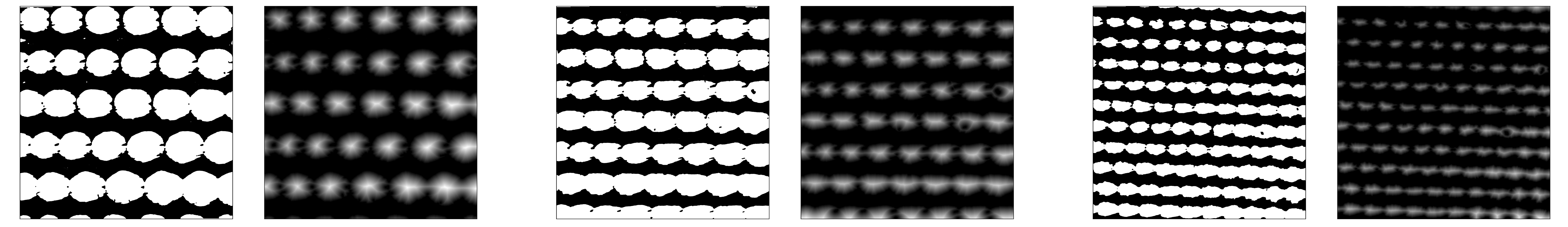}
\end{minipage}
\begin{minipage}[t]{0.32\textwidth}
\centering
(a) 0\% Overlap
\end{minipage}
\begin{minipage}[t]{0.32\textwidth}
\centering
(b) 25\% Overlap
\end{minipage}
\begin{minipage}[t]{0.32\textwidth}
\centering
(c) 50\% Overlap
\end{minipage}
\caption{Example binarized and distance transformed images for each overlap ratio. (a) shows the 0\% overlap image thresholded at a height of 0.47, (b) is the 25\% overlap image thresholded at 0.47 and (c) shows the 50\% overlap image thresholded at 0.51. Note: Binarized images are shown on the left and distance transformed images are shown on the right.}
\label{fig:thresh_ims}
\end{figure*}

\begin{figure*}[t]
\centering
\includegraphics[width=0.6\textwidth]{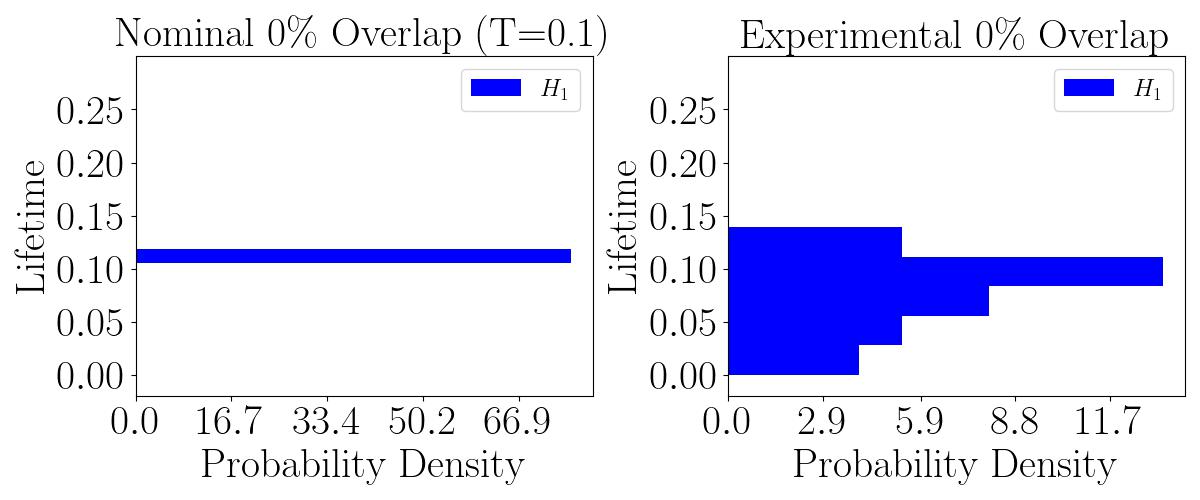} \caption{Roundness lifetime histogram example at a threshold height of 0.1 from the 0\% overlap image.}
\label{fig:hist_ex}
\end{figure*}

\begin{figure*}[t]
\centering
\includegraphics[width=0.9\textwidth]{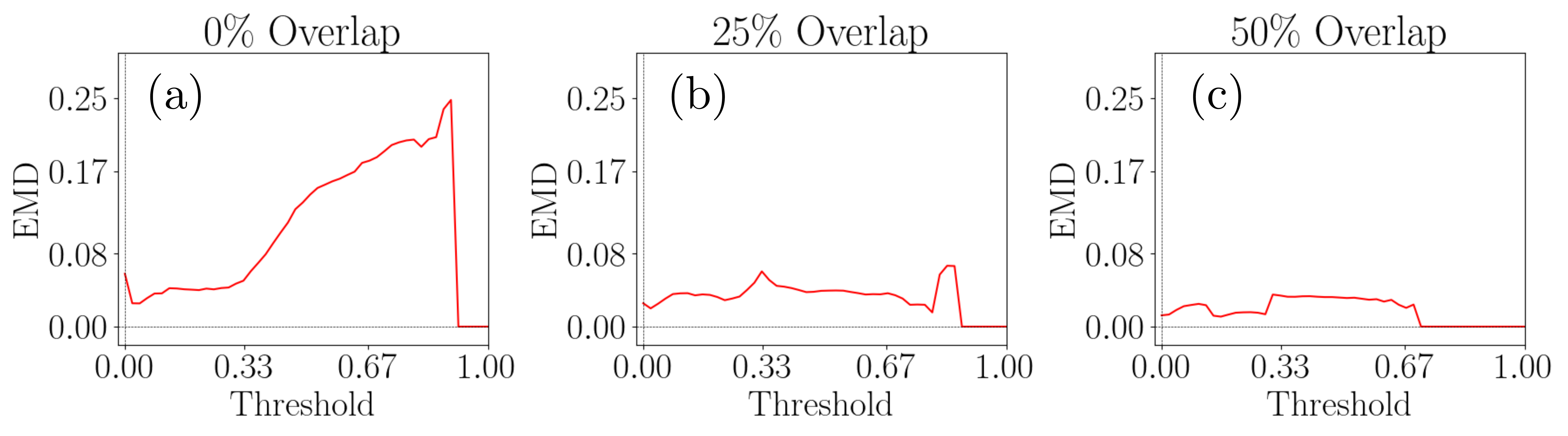} 
\caption{Earth movers distance between the experimental and theoretical roundness lifetime distributions as a function of threshold height for each overlap ratio.}
\label{fig:emd_plots}
\end{figure*}

Experimental images were thresholded and distance transformed at 50 heights ranging from 0 to 1 in the image and sublevel persistence was computed at each height. Figure~\ref{fig:thresh_ims} shows the thresholded and distance transformed images at various heights as an example.
The persistence lifetime histograms were used to compute the earth movers distance between the nominal and experimental distributions which provided a score at each height in the image and therefore information about the roundness over the entire depth of the strikes. Thresholding was started at the reference point ($T=0$) found from the filtered persistence diagrams. Histograms such as the ones in Fig.~\ref{fig:hist_ex} were generated at each height to visually compare the theoretical distribution of persistence lifetimes to the experimental distribution. This process resulted in an earth movers distance distribution with respect to threshold height as shown in Fig.~\ref{fig:emd_plots}. We expect the experimental distributions to be identical to the theoretical distributions and therefore have an earth movers distance of 0 at each height. Any deviation from 0 indicates a change in the uniformity of the roundness. The generalized roundness score was computed for each image by taking the area under the curves in Fig.~\ref{fig:emd_plots}. Qualitatively, we see that the 0\% image has the most deviation in the roundness when compared to the theoretical model due to its larger area under the earth movers distance curve. Similarly, the 50\% overlap image has the most consistent roundness due to its smaller area. To truly compare these plots, the scores need to be computed because the domain for each overlap ratio was different. The roundness scores were computed using Eq.~\eqref{eq:sphere_rd_score}
because the strikes were nominally spherical. The computed scores are shown in Table~\ref{tab:rd_scores} where a higher score corresponds to the roundness distribution being closer to nominal.

\begin{table}[htbp]
\caption{\label{tab:rd_scores}Computed roundness scores for each overlap ratio. Note that a higher score corresponds to the texture being closer to nominal.}
\centering
\begin{TAB}(r,0.1cm,0.1cm)[5pt]{|c|c|c|c|}{|c|c|}%
  Overlap Ratio & 0\% &25\%  & 50\%\ \\ 
   $\bar{R}$ & 30.82\% & 70.02\% & 74.26\% \\
\end{TAB}
\end{table}

As expected, the 50\% overlap image showed the highest roundness score indicating that this image had a more uniform roundness distribution, and the 0\% image had the lowest roundness score meaning it had the most deviation from nominal.

\section{Analysis Software}
The software used for this analysis was implemented as a texture analysis module in the \textit{teaspoon} python library for topological signal processing \cite{Munch2018}. This code provides functions for computing the depth and roundness scores between two input images and the user is responsible for supplying the nominal and experimental image arrays. The teaspoon functions utilize cubical ripser for sublevel persistence computations \cite{kaji2020cubical}.

To generate the nominal images, the process presented in Section~\ref{sec:verification} was followed using SolidWorks to model the surfaces and the algorithm in \cite{pc2im} to convert the CAD model to a point cloud that could be converted to an image.

\section{Acknowledgements}
This material is based upon work supported by the Air Force Office of Scientific Research, United States of America under award number FA9550-22-1-0007.

\section{Conclusion}
\label{sec:conclusion}

A novel approach to texture analysis was developed to describe specified features in a texture using topological data analysis. The tools were presented as an application to the surface treatment process piezo vibration striking treatment (PVST) in which a metal surface is impacted in a regular pattern by a tool on a CNC machine leaving a texture on the surface. Strike depth and roundness were successfully characterized using sublevel persistent homology, and scores were devised to quantify the features in the textural images relative to the nominal texture. In general, the higher overlap ratio images were found to provide more consistent strikes which could be due to the higher density of impacts on the surface. Two methods were also presented for generalizing the application of PVST of which the authors recommend using the CAD model method for an arbitrary tool shape. These tools allow for engineers to quantify specific features in a texture, a process which has typically been conducted qualitatively by manual inspection in the past. The scores obtained for depth and roundness features can be used to measure the effectiveness of the process that produced the pattern, and in future work, we plan to utilize these scores for extracting information on the material properties of the work piece.

\begin{appendices}
\section{Appendix  - Verifying Theoretical Results}

\label{sec:verification}

\begin{figure*}[htbp ]
	\centering
	\includegraphics[width=0.8\textwidth]{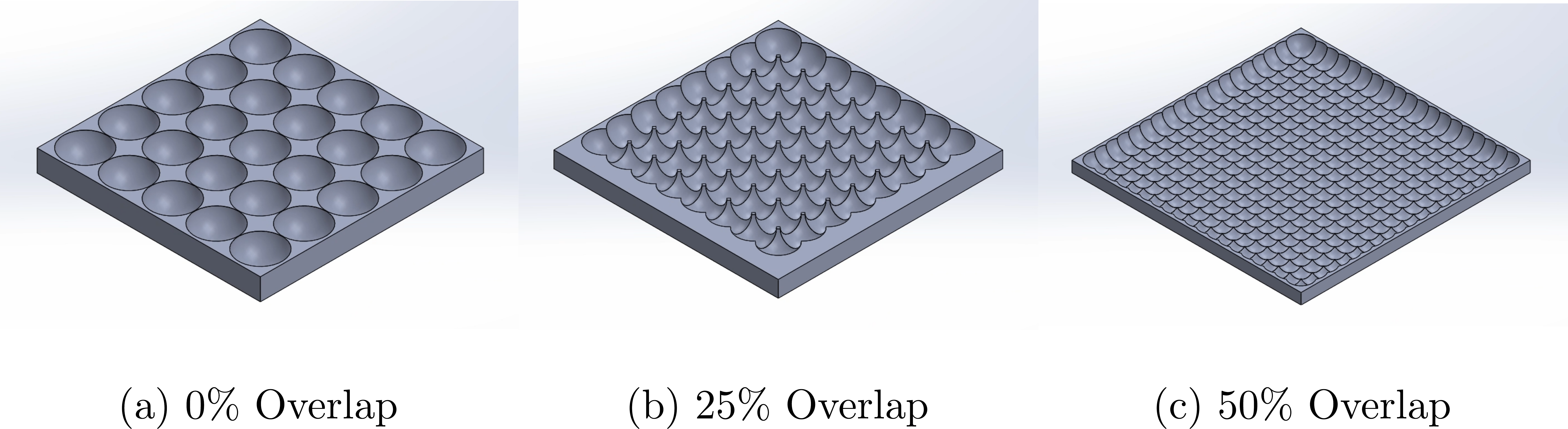}
	\caption{Ideal PVST grid CAD models at various overlap ratios. (a) 0\% overlap, (b) 25\% overlap, and (c) 50\% overlap.}
	\label{fig:cad_surfaces}
\end{figure*}

In order to verify the expressions in Section~\ref{sec:expressions_depth}, we manufactured gray scale images consisting of perfect PVST strikes in the expected patterns, and computed sublevel persistence to determine whether the results are consistent with the expressions. CAD models were created to model the expected surfaces for 0, 25, and 50\% overlap ratios as shown in Fig.~\ref{fig:cad_surfaces}. The number of strikes in each case was decided by assuming a 2.5 $\times$ 2.5 mm surface and a striking frequency of 100 Hz. Knowing these two parameters allowed for the in plane speeds to be set to obtain a specified overlap ratio. Note that the model was set up to only allow for full strikes and any fractional strike outside of the 2.5 $\times$ 2.5 mm window was ignored.

\begin{figure*}[t]
	\centering
	\includegraphics[width=0.8\textwidth]{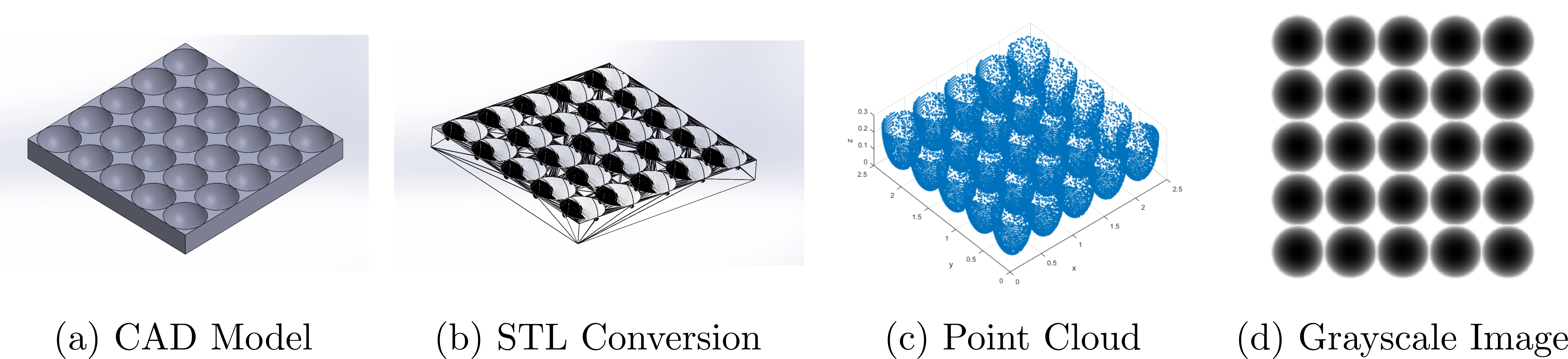}
	\caption{Pipeline for converting the PVST grid CAD model into a grayscale image. (a) shows the original CAD model, (b) the resulting STL file, (c) shows the point cloud obtained from the STL file, and (d) the final grayscale depth image.}
	\label{fig:im_pipeline}
\end{figure*}

To compute the sublevel persistence of a nominal texture the surface CAD model needed to be manipulated into a form that was compatible with the cubical ripser. The image pipeline shown in Fig.~\ref{fig:im_pipeline} was used to convert the CAD information into a gray scale image and subsequently a CSV file for cubical ripser. The CAD model was scaled up by a factor of 10000 to increase the resolution of the point cloud. This was necessary to mitigate the Solidworks STL resolution limitations, but the results were not affected due to the normalization of the points at a later step. 
A Matlab script was implemented to convert the high resolution STL files into point clouds. Note that the point cloud shown in Fig.~\ref{fig:im_pipeline}c was only plotting one point per 75 points for viewing clarity. After converting the model to a point cloud, the algorithm in \cite{pc2im} was used to convert the point cloud to a gray scale image and a bilinear interpolation created a smooth image as shown in Fig.~\ref{fig:im_pipeline}d. 

\subsection{Strike Depths}

\begin{figure*}[htbp]
\centering
\includegraphics[width=0.8\textwidth]{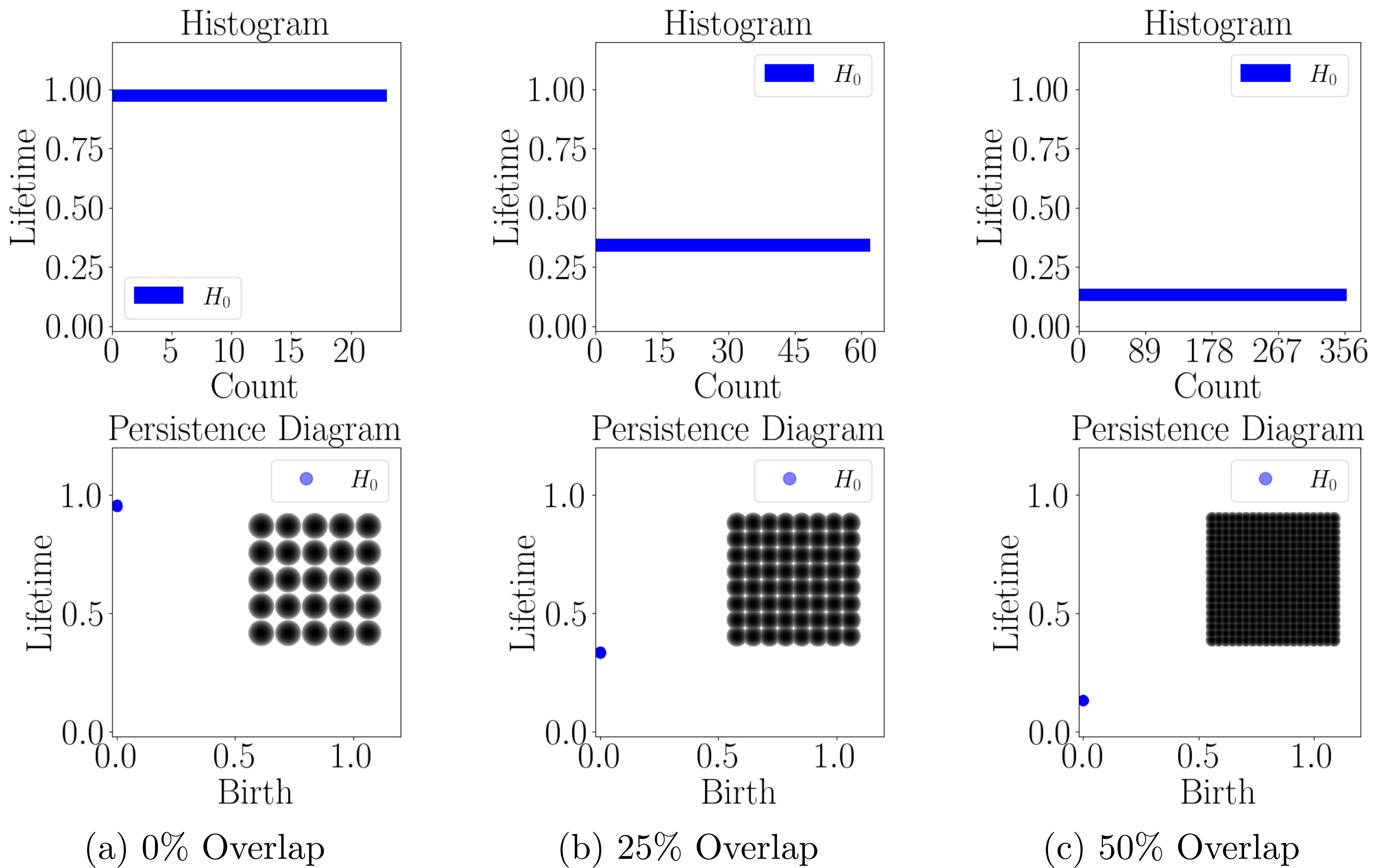}
\caption{Nominal CAD surface striking depth persistence diagrams and histograms for each overlap ratio.}
\label{fig:nom_pds}
\end{figure*}

This process was applied to the 0\%, 25\%, and 50\% overlap ratio grids and persistence diagrams were generated for each case as shown in Fig.~\ref{fig:nom_pds}.
Table~\ref{tab:valuecomparison} shows a comparison of the expected lifetimes using the derived results and the results obtained from the CAD model persistence. We see that the lifetimes obtained were exceedingly close to the expected results. The percent differences in each case being below 5\% allowed for the theoretical results for the striking depths to be verified and used to compare with the experimental images. 

\begin{table}[t]
\caption{\label{tab:valuecomparison}Comparison of the theoretical model lifetimes and the CAD Model generated striking \textbf{depth} lifetimes}
\centering
\begin{TAB}(r,0.1cm,0.1cm)[5pt]{|c|c|c|c|}{|c|c|c|c|}%
  Overlap Ratio & 0\% & 25\%  & 50\%\ \\ 
   Theoretical Lifetime ($\overline{h}$) & $1$ & $0.339$ & $0.134$ \\
   CAD Model Lifetime & $0.954$ & $0.337$ & $0.1337$ \\
  \textbf{Percent Difference} & \textbf{4.6\%} & \textbf{0.5\%} & \textbf{0.22\%} \\
\end{TAB}
\end{table}

\subsection{Strike Roundness}

To test the theoretical results for the strike roundness, we threshold the images at two different heights. One height below the critical height and one above to determine if both results are consistent with the expressions.

\subsubsection{Roundness - No Overlap}
\begin{figure*}[htbp]
	\centering
	\includegraphics[width=0.8\textwidth]{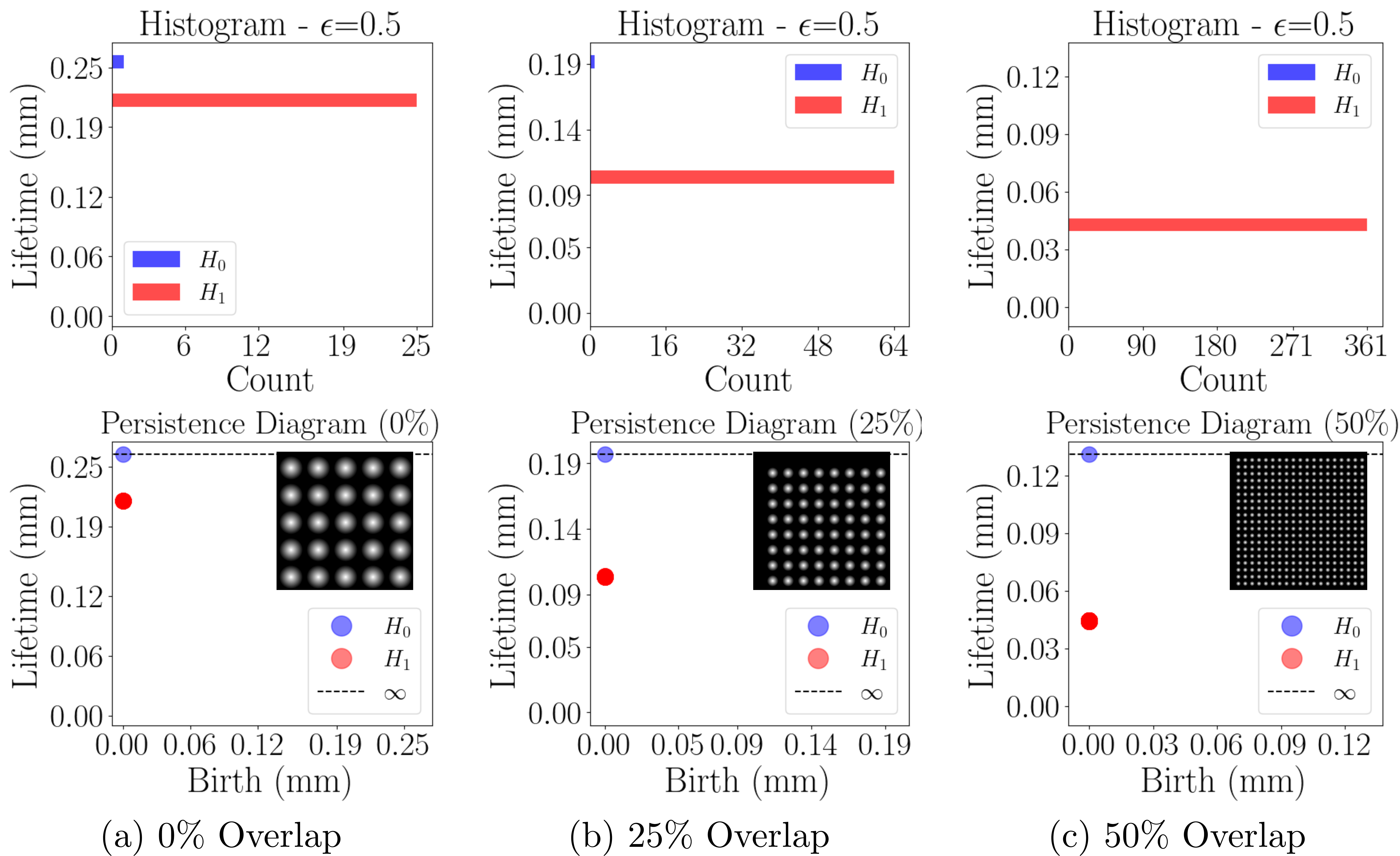}
	\caption{CAD model distance transformed image (strike roundness) persistence diagrams for $\epsilon = 0.5$ ($T<\overline{h}$) at each overlap ratio.}
	\label{fig:half_pds}
\end{figure*}

First, the images were thresholded at half of the nominal depth ($\epsilon = 0.5$)
\begin{equation}
	T = 0.5 h,
\end{equation}
where $T$ is the image threshold height and anything above $T$ is set to black and any pixel below $T$ is set to white. The threshold and distance transform results for the nominal images are shown in Fig.~\ref{fig:half_pds}. Because half of the nominal depth was chosen for thresholding, we expect zero overlap in each case. This means that the persistence diagram should have $n^2$ 1D classes born at zero that die at $\sigma = \sqrt{(2R-T)T}$. The corresponding persistence diagrams for the half nominal depth threshold are shown in Fig.~\ref{fig:half_pds}.
We see that the 1D persistence shows the expected number of loops that are born at time 0 and die at various heights. Using Eq.~\eqref{eq:distToPix}, we have converted the pixel distances into distances in mm using $W=2.55$~mm and $P=5000$.
\begin{table}[ht]
\caption{\label{tab:measured_radius50}Comparison of CAD model persistence results and theoretical strike \textbf{roundness} for each overlap ratio ($\epsilon=0.5$).}
\centering
\begin{TAB}(r,0.1cm,0.1cm)[5pt]{|c|c|c|c|}{|c|c|c|c|}%
  Overlap Ratio & 0\% & 25\%  & 50\%\ \\ 
   1-D Death [mm] & 0.2157 & 0.1035 & 0.0445 \\
   $\sigma$ [mm] & 0.21651 & 0.10438 & 0.04498 \\
   \textbf{Percent Difference} & \textbf{0.372\%} & \textbf{0.843\%} & \textbf{1.068\%} \\
\end{TAB}
\end{table}
The results in Table~\ref{tab:measured_radius50} are exceedingly close to the expected values from the theory. This implies that the theoretical model is correct for the case when the circles are not touching. 

\subsubsection{Roundness - Overlap}

\begin{figure*}[htbp]
	\centering
	\includegraphics[width=\textwidth]{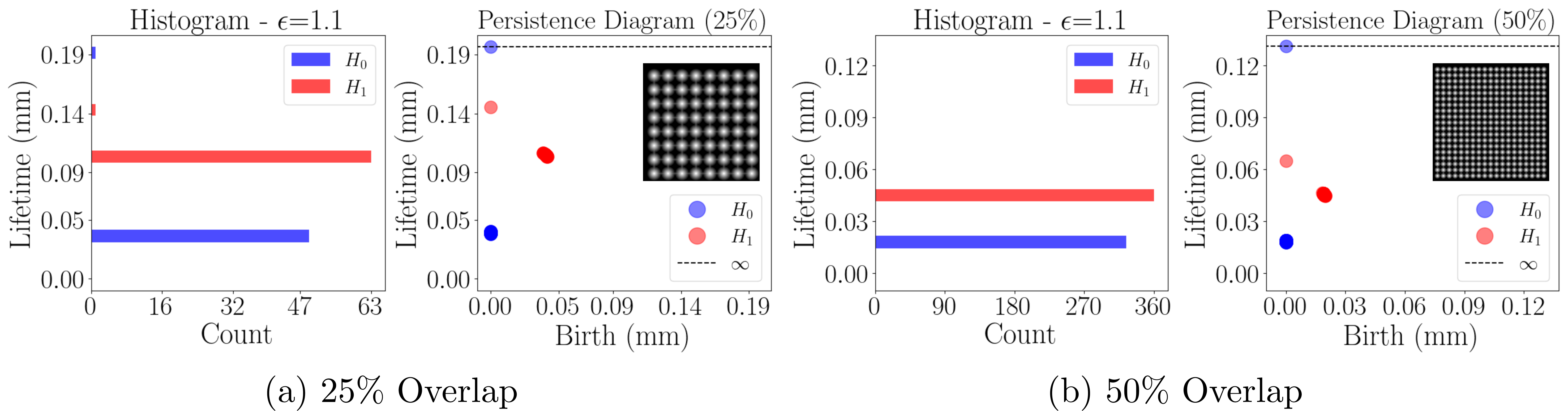}
	\caption{CAD model distance transformed image (strike roundness) persistence diagrams for $\epsilon = 1.1$ ($T>\overline{h}$) at each overlap ratio.}
	\label{fig:pds_110}
\end{figure*}

To consider a case of overlapping circles, only the 25\% and 50\% images can be considered. We test the theory for images that contain overlap by computing persistence on the CAD models at $\epsilon=1.1$. Figure~\ref{fig:pds_110} shows the thresholded images at this height.
The corresponding persistence diagrams for $\epsilon=1.1$ are also shown in Fig.~\ref{fig:pds_110}. We see that there are $n^2-1$ 1D classes born around $a$ that die at the radius $\sigma$. The experimental values are compared to the nominal values in Table~\ref{tab:measured_radius110}. It was clear that the results were nearly identical to the theoretical results which verifies the expressions from Section~\ref{sec:roundness_expressions}.
\begin{table}[H]
\caption{\label{tab:measured_radius110}Comparison of CAD model persistence results and theoretical strike \textbf{roundness} for each overlap ratio ($\epsilon=1.1$).}
\centering
\begin{TAB}(r,0.1cm,0.1cm)[5pt]{|c|c|c|}{|c|c|c|c|c|c|c|}%
  Overlap Ratio & 25\%  & 50\%\ \\ 
   1-D Birth [mm] & 0.0405 & 0.01916  \\
   $a$ [mm] &      0.03917 & 0.01897  \\
   \textbf{Percent Difference} & \textbf{3.396\%} & \textbf{1.014\%} \\
   1-D Death [mm] & 0.1452 & 0.06480  \\
   $\sigma$ [mm] & 0.1459 & 0.0653  \\
   \textbf{Percent Difference} & \textbf{0.533\%} & \textbf{0.788\%}  \\
\end{TAB}
\end{table}

\section{Appendix  - Score Noise Study}

\label{sec:noise}

\subsection{Feature Depth}
\begin{figure*}[htbp]
	\centering
	\begin{minipage}[t]{0.49\textwidth}
		\centering
		\includegraphics[width=\textwidth]{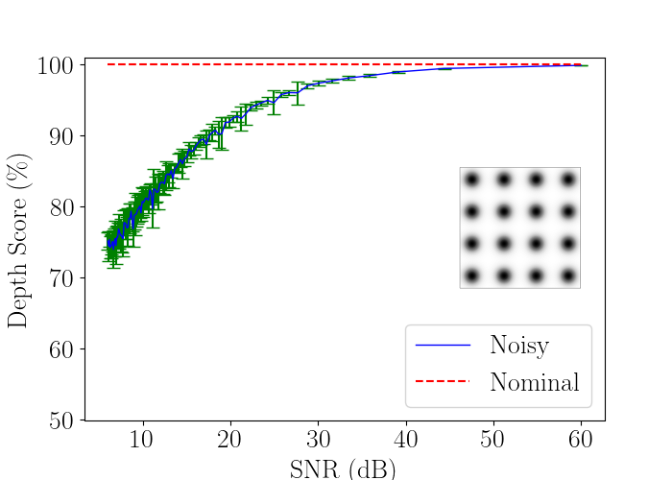}
		(a) Depth
	\end{minipage}
	\begin{minipage}[t]{0.49\textwidth}
		\centering
		\includegraphics[width=\textwidth]{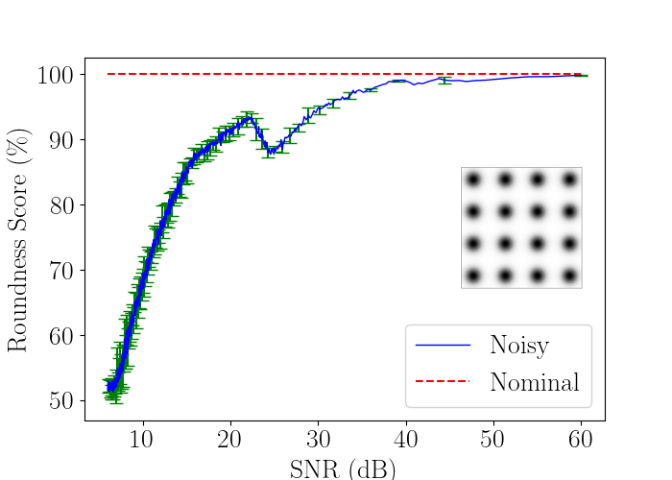}
		(b) Roundness
	\end{minipage}
	\caption{Texture quantification scores plotted as a function of the SNR in dB. The average score of 10 trials is plotted as a solid line and the dashed line indicates the true score of the feature depths. Error bars are shown at one standard deviation of the 10 trials at each SNR.}
	\label{fig:noise_study}
\end{figure*}

A noise study was conducted on the feature depth score by generating a synthetic texture using a superposition of two-dimensional Gaussian distributions in a four by four grid as the features. The synthetic surface is shown in Fig.~\ref{fig:noise_study}(a). Gaussian noise was added to this image by specifying an amplitude on a normal distribution and comparing this amplitude to the nominal strike depth (1) to generate a signal to noise ratio (SNR) in dB. The depth score was then computed and plotted over a range of SNRs to quantify the noise robustness of the score. Ten trials were conducted at each SNR with the average score plotted with error bars indicating one standard deviation from the mean. The resulting plot for the depth score is shown in Fig.~\ref{fig:noise_study}(a). We see that the depth score remains within 5\% of the nominal score (100\%) for SNRs down to approximately 25 dB. 

\subsection{Feature Roundness}
The roundness score was then computed with varying SNR in the synthetic surface using same process and synthetic surface. Ten trials were conducted for each SNR and the average roundness score was plotted as a function of SNR with error bars indicating one standard deviation from the average shown in Fig.~\ref{fig:noise_study}(b). 
We see that the roundness score remains within 5\% of the nominal score down to approximately 30 dB. It is also clear that the variability in the roundness score is smaller compared to the depth score. This was likely due to each roundness score being made up of 30 earth movers distance computations which reduces the effect of outliers because a single outlier in the earth movers distance plot will not have a significant effect on the area under the curve.

\end{appendices}

\begin{appendices}
\section{Appendix  - Estimating Surface Slope and Angularity}

\label{sec:slopes}

During the PVST process, the tool is set to strike the same depth for each cycle. If the sample surface is not perfectly flat relative to the CNC datum, the strike depths will vary across the surface. We see in Fig.~\ref{fig:pd_filtering}(c) that the strikes toward the bottom right of the image are deeper in general compared to the opposite corner due to the larger birth times in the top left corner. As a result, we expect that the surface is sloped toward the bottom right corner and we can approximate this slope by fitting a regression plane to the point cloud shown in Fig.~\ref{fig:pd_filtering}(c). For this image, the resulting plane has the form, 
\begin{equation}
	z = 0.1508 - 0.0003315 i_x - 0.0001764 i_y
\end{equation}
where $i_x$ and $i_y$ are the pixel indices in the $x$ and $y$ directions respectively and $z$ is the height in the image. The slope coefficients on the $i_x$ and $i_y$ terms have units of $\frac{1}{\textrm{pixel}}$ due to the normalization of the depths in the image. The slopes can be converted to units of $\frac{\mu m}{mm}$ using the maximum depth of the image in microns, the width of the image in millimeters and the number of pixels along one axis in the image. The resulting slopes are, $m_x=-0.941$ and $m_y=-0.501$ $\frac{\mu m}{mm}$. In other words, for each millimeter increase in the horizontal direction in this image, we expect the strike depth to increase by about 0.941 microns, This increase corresponds with the top of the surface in this location being closer to the CNC tool. This means that the sample is sloped in the opposite directions. These slopes helped explain why the observed strike depths are deeper toward the bottom right corner of the image and why the image thresholding cannot provide an ideal quantification of the roundness of the strikes and persistence diagram filtering needed to be used to get an optimal reference height. The slopes for the 25 and 50\% overlap images are shown in Table~\ref{tab:slopes}. Note that the image coordinate system was used for these slopes so the negative $y$ direction points toward the top of the image.

\begin{table}[H]
\caption{\label{tab:slopes}Measured Radius at Half Nominal Depth}
\centering
\begin{TAB}(r,0.1cm,0.1cm)[5pt]{|c|c|c|}{|c|c|c|c|}%
  Direction & $m_x$ [$\mu m/mm$]  & $m_y$ [$\mu m/mm$]\ \\ 
   0\% Overlap & $-0.941$ & $-0.501$  \\
   25\% Overlap &      $-1.211$ & $-2.307$  \\
   50\% Overlap & $-0.832$ & $-0.475$ \\
\end{TAB}
\end{table}
\end{appendices}

\bibliographystyle{ieeetr}

\bibliography{Sections/bibliography}

\begin{thebibliography}{10}

\bibitem{Masokano2020}
I.~B. Masokano, W.~Liu, S.~Xie, D.~F.~H. Marcellin, Y.~Pei, and W.~Li, ``The
  application of texture quantification in hepatocellular carcinoma using ct
  and mri: a review of perspectives and challenges,'' {\em Cancer Imaging},
  2020.

\bibitem{Ymeti2017}
I.~Ymeti, D.~Shrestha, V.~Jetten, C.~Lievens, and and, ``Using color, texture
  and object-based image analysis of multi-temporal camera data to monitor soil
  aggregate breakdown,'' {\em Sensors}, vol.~17, p.~1241, may 2017.

\bibitem{Fan2019}
F.~Gao and Y.~Lu, ``Moving target detection using inter-frame difference
  methods combined with texture features and lab color space,'' in {\em 2019
  International Conference on Artificial Intelligence and Advanced
  Manufacturing (AIAM)}, pp.~76--81, 2019.

\bibitem{Thomas1998}
T.~Thomas, ``Trends in surface roughness,'' {\em International Journal of
  Machine Tools and Manufacture}, vol.~38, pp.~405--411, may 1998.

\bibitem{Spierings2013}
A.~Spierings, T.~Starr, and K.~Wegener, ``Fatigue performance of additive
  manufactured metallic parts,'' {\em Rapid Prototyping Journal}, vol.~19,
  pp.~88--94, 3 2013.

\bibitem{Frazier2014}
W.~E. Frazier, ``Metal additive manufacturing: A review,'' {\em Journal of
  Materials Engineering and Performance}, vol.~23, pp.~1917--1928, apr 2014.

\bibitem{Chan2012}
K.~S. Chan, M.~Koike, R.~L. Mason, and T.~Okabe, ``Fatigue life of titanium
  alloys fabricated by additive layer manufacturing techniques for dental
  implants,'' {\em Metallurgical and Materials Transactions A}, vol.~44,
  pp.~1010--1022, oct 2012.

\bibitem{Yin2003}
H.~Yin and T.~Emi, ``Marangoni flow at the gas/melt interface of steel,'' {\em
  Metallurgical and Materials Transactions B}, vol.~34, pp.~483--493, 10 2003.

\bibitem{Gu2009}
D.~Gu and Y.~Shen, ``Balling phenomena in direct laser sintering of stainless
  steel powder: Metallurgical mechanisms and control methods,'' {\em Materials
  {\&} Design}, vol.~30, pp.~2903--2910, sep 2009.

\bibitem{Gu2015}
D.~Gu, {\em Laser Additive Manufacturing of High-Performance Materials}.
\newblock Springer Berlin Heidelberg, 2015.

\bibitem{LIU2022108068}
Y.~Liu, L.~Guo, H.~Gao, Z.~You, Y.~Ye, and B.~Zhang, ``Machine vision based
  condition monitoring and fault diagnosis of machine tools using information
  from machined surface texture: A review,'' {\em Mechanical Systems and Signal
  Processing}, vol.~164, p.~108068, 2022.

\bibitem{Khalifa2006}
O.~O. Khalifa, A.~Densibali, and W.~Faris, ``Image processing for chatter
  identification in machining processes,'' {\em The International Journal of
  Advanced Manufacturing Technology}, vol.~31, pp.~443--449, feb 2006.

\bibitem{Lei2016}
N.~Lei and M.~Soshi, ``Vision-based system for chatter identification and
  process optimization in high-speed milling,'' {\em The International Journal
  of Advanced Manufacturing Technology}, vol.~89, pp.~2757--2769, dec 2016.

\bibitem{Szydlowski2011}
M.~Szyd{\l}owski and B.~Powa{\l}ka, ``Chatter detection algorithm based on
  machine vision,'' {\em The International Journal of Advanced Manufacturing
  Technology}, vol.~62, pp.~517--528, dec 2011.

\bibitem{Li2020}
D.-D. Li, W.-M. Zhang, Y.-S. Li, F.~Xue, and J.~Fleischer, ``Chatter
  identification of thin-walled parts for intelligent manufacturing based on
  multi-signal processing,'' {\em Advances in Manufacturing}, vol.~9,
  pp.~22--33, apr 2020.

\bibitem{Tran2021}
M.-Q. Tran, M.~Elsisi, and M.-K. Liu, ``Effective feature selection with fuzzy
  entropy and similarity classifier for chatter vibration diagnosis,'' {\em
  Measurement}, vol.~184, p.~109962, nov 2021.

\bibitem{Zhu2020}
W.~Zhu, J.~Zhuang, B.~Guo, W.~Teng, and F.~Wu, ``An optimized convolutional
  neural network for chatter detection in the milling of thin-walled parts,''
  {\em The International Journal of Advanced Manufacturing Technology},
  vol.~106, pp.~3881--3895, jan 2020.

\bibitem{Bhat2016}
N.~N. Bhat, S.~Dutta, S.~K. Pal, and S.~Pal, ``Tool condition classification in
  turning process using hidden markov model based on texture analysis of
  machined surface images,'' {\em Measurement}, vol.~90, pp.~500--509, aug
  2016.

\bibitem{Bradley2001}
C.~Bradley and Y.~Wong, ``Surface texture indicators of tool wear - a machine
  vision approach,'' {\em The International Journal of Advanced Manufacturing
  Technology}, vol.~17, pp.~435--443, apr 2001.

\bibitem{Datta2013}
A.~Datta, S.~Dutta, S.~Pal, and R.~Sen, ``Progressive cutting tool wear
  detection from machined surface images using voronoi tessellation method,''
  {\em Journal of Materials Processing Technology}, vol.~213, pp.~2339--2349,
  dec 2013.

\bibitem{Li2016}
L.~Li and Q.~An, ``An in-depth study of tool wear monitoring technique based on
  image segmentation and texture analysis,'' {\em Measurement}, vol.~79,
  pp.~44--52, feb 2016.

\bibitem{Kerr2005}
D.~Kerr, J.~Pengilley, and R.~Garwood, ``Assessment and visualisation of
  machine tool wear using computer vision,'' {\em The International Journal of
  Advanced Manufacturing Technology}, vol.~28, pp.~781--791, may 2005.

\bibitem{Danesh2015}
M.~Danesh and K.~Khalili, ``Determination of tool wear in turning process using
  undecimated wavelet transform and textural features,'' {\em Procedia
  Technology}, vol.~19, pp.~98--105, 2015.

\bibitem{Kassim2004}
A.~Kassim, Z.~Mian, and M.~Mannan, ``Connectivity oriented fast hough transform
  for tool wear monitoring,'' {\em Pattern Recognition}, vol.~37,
  pp.~1925--1933, sep 2004.

\bibitem{Zhu2017}
K.~Zhu and X.~Yu, ``The monitoring of micro milling tool wear conditions by
  wear area estimation,'' {\em Mechanical Systems and Signal Processing},
  vol.~93, pp.~80--91, sep 2017.

\bibitem{stkepien2014research}
K.~St{\c{e}}pie{\'n}, ``Research on a surface texture analysis by digital
  signal processing methods,'' {\em Tehnicki Vjesnik-Technical Gazette},
  vol.~21, no.~3, pp.~485--493, 2014.

\bibitem{Santiago2011}
A.~J.~S. Santiago, A.~J. Yuste, J.~E.~M. Exp{\'{o}}sito, S.~G. Gal{\'{a}}n,
  R.~P. Prado, J.~M. Maqueira, and S.~Bruque, ``Real-time image texture
  analysis in quality management using grid computing: an application to the
  {MDF} manufacturing industry,'' {\em The International Journal of Advanced
  Manufacturing Technology}, vol.~58, pp.~1217--1225, aug 2011.

\bibitem{xie2008review}
X.~Xie, ``A review of recent advances in surface defect detection using texture
  analysis techniques,'' {\em ELCVIA: electronic letters on computer vision and
  image analysis}, pp.~1--22, 2008.

\bibitem{ozturk2015comparison}
{\c{S}}.~{\"O}zt{\"u}rk and B.~Akdemir, ``Comparison of edge detection
  algorithms for texture analysis on glass production,'' {\em Procedia-Social
  and Behavioral Sciences}, vol.~195, pp.~2675--2682, 2015.

\bibitem{Vijaykumar2015}
V.~R. Vijaykumar and S.~Sangamithirai, ``Rail defect detection using gabor
  filters with texture analysis,'' {\em 2015 3rd International Conference on
  Signal Processing, Communication and Networking ({ICSCN})}, mar 2015.

\bibitem{Kilic2006}
M.~Kilic, S.~Hiziroglu, and E.~Burdurlu, ``Effect of machining on surface
  roughness of wood,'' {\em Building and Environment}, vol.~41, pp.~1074--1078,
  aug 2006.

\bibitem{Myshkin2003}
N.~Myshkin, A.~Grigoriev, S.~Chizhik, K.~Choi, and M.~Petrokovets, ``Surface
  roughness and texture analysis in microscale,'' {\em Wear}, vol.~254,
  pp.~1001--1009, jul 2003.

\bibitem{Josso2002}
B.~Josso, D.~R. Burton, and M.~J. Lalor, ``Frequency normalised wavelet
  transform for surface roughness analysis and characterisation,'' {\em Wear},
  vol.~252, pp.~491--500, mar 2002.

\bibitem{AlMangour2016}
B.~AlMangour and J.-M. Yang, ``Improving the surface quality and mechanical
  properties by shot-peening of 17-4 stainless steel fabricated by additive
  manufacturing,'' {\em Materials \&$;$ Design}, vol.~110, pp.~914--924, nov
  2016.

\bibitem{Hatamleh2008}
O.~Hatamleh, ``The effects of laser peening and shot peening on mechanical
  properties in friction stir welded 7075-t7351 aluminum,'' {\em Journal of
  Materials Engineering and Performance}, vol.~17, pp.~688--694, oct 2008.

\bibitem{Liu2020}
Y.~Liu, Y.~Cao, H.~Zhou, X.~Chen, Y.~Liu, L.~Xiao, X.~Huan, Y.~Zhao, and
  Y.~Zhu, ``Mechanical properties and microstructures of commercial-purity
  aluminum processed by rotational accelerated shot peening plus cold
  rolling,'' {\em Advanced Engineering Materials}, vol.~22, no.~1, p.~1900478,
  2020.

\bibitem{Maleki2019}
E.~Maleki and O.~Unal, ``Shot peening process effects on metallurgical and
  mechanical properties of 316 l steel via: Experimental and neural network
  modeling,'' {\em Metals and Materials International}, vol.~27, pp.~262--276,
  sep 2019.

\bibitem{Jamalian2019}
M.~Jamalian and D.~P. Field, ``Effects of shot peening parameters on gradient
  microstructure and mechanical properties of {TRC} {AZ}31,'' {\em Materials
  Characterization}, vol.~148, pp.~9--16, feb 2019.

\bibitem{Xie2016}
L.~Xie, Y.~Wen, K.~Zhan, L.~Wang, C.~Jiang, and V.~Ji, ``Characterization on
  surface mechanical properties of ti{\textendash}6al{\textendash}4v after shot
  peening,'' {\em Journal of Alloys and Compounds}, vol.~666, pp.~65--70, may
  2016.

\bibitem{Guo2013}
P.~Guo and K.~F. Ehmann, ``An analysis of the surface generation mechanics of
  the elliptical vibration texturing process,'' {\em International Journal of
  Machine Tools and Manufacture}, vol.~64, pp.~85--95, jan 2013.

\bibitem{Kurniawan2016}
R.~Kurniawan, G.~Kiswanto, and T.~J. Ko, ``Micro-dimple pattern process and
  orthogonal cutting force analysis of elliptical vibration texturing,'' {\em
  International Journal of Machine Tools and Manufacture}, vol.~106,
  pp.~127--140, jul 2016.

\bibitem{Jiang2020}
J.~Jiang, S.~Sun, D.~Wang, Y.~Yang, and X.~Liu, ``Surface texture formation
  mechanism based on the ultrasonic vibration-assisted grinding process,'' {\em
  International Journal of Machine Tools and Manufacture}, vol.~156, p.~103595,
  sep 2020.

\bibitem{chen2021force}
J.~Chen, Y.~Xu, J.~Sandoval, P.~Kwon, and Y.~Guo, ``On force-displacement
  characteristics and surface deformation in piezo vibration striking treatment
  (pvst),'' {\em Journal of Manufacturing Science and Engineering}, pp.~1--27,
  2021.

\bibitem{Bharati2004}
M.~H. Bharati, J.~Liu, and J.~F. MacGregor, ``Image texture analysis: methods
  and comparisons,'' {\em Chemometrics and Intelligent Laboratory Systems},
  vol.~72, pp.~57--71, jun 2004.

\bibitem{srinivasan2008statistical}
G.~Srinivasan and G.~Shobha, ``Statistical texture analysis,'' in {\em
  Proceedings of world academy of science, engineering and technology},
  vol.~36, pp.~1264--1269, 2008.

\bibitem{Materka1998}
A.~Materka, M.~Strzelecki, {\em et~al.}, ``Texture analysis methods--a
  review,'' {\em Technical university of lodz, institute of electronics, COST
  B11 report, Brussels}, vol.~10, no.~1.97, p.~4968, 1998.

\bibitem{Wang2008}
Z.-Z. Wang and J.-H. Yong, ``Texture analysis and classification with linear
  regression model based on wavelet transform,'' {\em IEEE transactions on
  image processing}, vol.~17, no.~8, pp.~1421--1430, 2008.

\bibitem{Motta2018}
F.~C. Motta, R.~Neville, P.~D. Shipman, D.~A. Pearson, and R.~M. Bradley,
  ``Measures of order for nearly hexagonal lattices,'' {\em Physica D:
  Nonlinear Phenomena}, vol.~380-381, pp.~17--30, oct 2018.

\bibitem{Yesilli2022}
M.~C. Yesilli, M.~M. Chumley, J.~Chen, F.~A. Khasawneh, and Y.~Guo,
  ``{Exploring Surface Texture Quantification in Piezo Vibration Striking
  Treatment (PVST) Using Topological Measures},'' {\em International
  Manufacturing Science and Engineering Conference}, vol.~Volume 2:
  Manufacturing Processes; Manufacturing Systems, 06 2022.
\newblock V002T05A061.

\bibitem{dey2022computational}
T.~K. Dey and Y.~Wang, {\em Computational topology for data analysis}.
\newblock Cambridge University Press, 2022.

\bibitem{Kaji2020}
S.~Kaji, T.~Sudo, and K.~Ahara, ``Cubical ripser: Software for computing
  persistent homology of image and volume data,'' 2020.

\bibitem{CohenSteiner2006}
D.~Cohen-Steiner, H.~Edelsbrunner, and J.~Harer, ``Stability of persistence
  diagrams,'' {\em Discrete {\&} Computational Geometry}, vol.~37,
  pp.~103--120, dec 2006.

\bibitem{rices_rule}
D.~M. Lane, ``Online statistics education.'' \url{http://onlinestatbook.com/},
  4 2013.

\bibitem{EMD}
SciPy, ``Wasserstein distance.''
  \url{https://docs.scipy.org/doc/scipy/reference/generated/scipy.stats.wasserstein_distance.html},
  2008.

\bibitem{Arizmendi2019}
M.~Arizmendi, A.~Jim{\'{e}}nez, W.~E. Cumbicus, M.~Estrems, and M.~Artano,
  ``Modelling of elliptical dimples generated by five-axis milling for surface
  texturing,'' {\em International Journal of Machine Tools and Manufacture},
  vol.~137, pp.~79--95, feb 2019.

\bibitem{Grob1998}
C.~Grob and T.-K. Strempel, ``On generalizations of conics and on a
  generalization of the fermat- torricelli problem,'' {\em The American
  Mathematical Monthly}, vol.~105, p.~732, oct 1998.

\bibitem{Munch2018}
E.~Munch, ``Teaspoon.'' \url{https://github.com/lizliz/teaspoon}, 2018.

\bibitem{kaji2020cubical}
S.~Kaji, T.~Sudo, and K.~Ahara, ``Cubical ripser: Software for computing
  persistent homology of image and volume data,'' 2020.

\bibitem{pc2im}
V.~Behravan, ``pointcloud2image( x,y,z,numr,numc ).''
  \url{https://www.mathworks.com/matlabcentral/fileexchange/55031-pointcloud2image-x-y-z-numr-numc},
  1 2016.

\end{thebibliography}

\end{document}